%% file: main.tex
\documentclass[conference]{IEEEtran}
\IEEEoverridecommandlockouts

\usepackage{cite}
\usepackage[T1]{fontenc}
\usepackage{amsmath,amssymb,amsfonts}
\usepackage{algorithmic}
\usepackage{graphicx}
\usepackage{textcomp}
\usepackage{xcolor}

\def\BibTeX{{\rm B\kern-.05em{\sc i\kern-.025em b}\kern-.08em
    T\kern-.1667em\lower.7ex\hbox{E}\kern-.125emX}}
\begin{document}

\title{NOVA: Technology-Architecture Co-Design of Near-Memory Processing for Attention-SSM-MoE Hybrid LLM Inference\\
\thanks{\textsuperscript{*}Equal contributions}
\thanks{\copyright~2026 IEEE. Personal use of this material is permitted.
Permission from IEEE must be obtained for all other uses, in any current or
future media, including reprinting/republishing this material for advertising
or promotional purposes, creating new collective works, for resale or
redistribution to servers or lists, or reuse of any copyrighted component of
this work in other works. Accepted to the 59th IEEE/ACM International
Symposium on Microarchitecture (MICRO), 2026.}
}

\author{\IEEEauthorblockN{In-Jun Jung\textsuperscript{*}}
\IEEEauthorblockA{\textit{KAIST} \\
Daejeon, South Korea \\
injun@kaist.ac.kr}
\and
\IEEEauthorblockN{Jaeha Min\textsuperscript{*}}
\IEEEauthorblockA{\textit{KAIST} \\
Daejeon, South Korea \\
derekmin0807@kaist.ac.kr}
\and
\IEEEauthorblockN{Joo-Young Kim}
\IEEEauthorblockA{\textit{KAIST} \\
Daejeon, South Korea \\
jooyoung1203@kaist.ac.kr}
}

\maketitle

\begin{abstract}
 The rapid evolution of hybrid large language models (LLMs), which interleave grouped-query-attention (GQA), state-space model (SSM), and Mixture-of-Experts (MoE) layers, introduces two fundamental challenges for near-memory processing (NMP) architectures. First, the Technology Wall: the conventional $\boldsymbol{6F^2}$ DRAM cell is approaching its physical scaling limits at current 10nm-class nodes, making it difficult to meet the rapidly increasing memory capacity demands driven by MoE models with hundreds of experts. Second, the Architecture Wall: existing NMP designs are optimized for narrow arithmetic intensity (Op/B) ranges and cannot efficiently support the highly heterogeneous compute characteristics of hybrid LLMs, spanning memory-bound SSM layers, compute-intensive GQA layers, and large Op/B variations across experts.

In this paper, we propose NOVA, a technology-architecture co-designed NMP system that simultaneously overcomes both walls. On the technology side, NOVA combines a $\boldsymbol{4F^2}$ vertical channel transistor (VCT) DRAM cell with a peri-over-cell (POC) structure to achieve approximately 2$\times$ memory density at iso-area compared to conventional $\boldsymbol{6F^2}$-based DRAM, enabling continued scaling into sub-10nm nodes. On the architecture side, NOVA repurposes the POC peripheral-die (peri-die) to host processing units, forming a 2-tier NMP architecture: Tier-1 (peri-die NMP), optimized for low-to-mid Op/B operations, and Tier-2 (base-die NMP), optimized for mid-to-high Op/B operations. By enabling parallel execution across tiers, NOVA effectively supports diverse compute patterns for hybrid LLMs, maximizing inference performance. Evaluated on state-of-the-art hybrid and MoE LLMs, including Nemotron3-Nano, Nemotron3-Super, Falcon-H1R, and Qwen3, NOVA achieves an average of 4.5$\times$ higher throughput, 69.8\% lower end-to-end latency, and 5$\times$ better energy efficiency compared to a GPU baseline, with only 3.9\% area overhead and no loss in memory capacity. 
\end{abstract}


\input{Outline/1_introduction}

\input{Outline/2_background}
\input{Outline/3_motivation}

\input{Outline/4_NOVA_arch}

\input{Outline/5_NOVA_mapping}

\input{Outline/6_evaluation}

\input{Outline/7_related_works}

\input{Outline/8_conclusion}

\section*{Acknowledgment}
This work was supported by the Institute of Information \& Communications Technology Planning \& Evaluation (IITP) grants funded by the Korea government (MSIT) (No. RS-2025-02264029, Integration and Validation of an AI Semiconductor-Based Data Center Training and Inference System, and No. IITP-2025-RS-2023-00256472, Graduate School of Artificial Intelligence Semiconductor) and by Samsung Electronics Co., Ltd (IO251211-14291-01).

\bibliographystyle{IEEEtran}
\bibliography{References}

\end{document}

%% file: Outline/1_introduction.tex
\section{Introduction}\label{Intro}
In modern large language model (LLM) inference systems, memory bandwidth and capacity have emerged as the primary bottlenecks determining both performance and cost-efficiency~\cite{ma2026challenges}. The widespread adoption of long-context inference~\cite{gemini1_5_long_ctx,llama3}, multimodal processing~\cite{openai2024gpt4technicalreport,geminiteam2025geminifamilyhighlycapable}, and reasoning~\cite{openai2024openaio1card,deepseek_r1} has driven a rapid increase in effective sequence lengths, thereby proportionally amplifying both the memory bandwidth demands and the capacity requirements of the key-value (KV) cache. Simultaneously, batching for high throughput proportionally increases the total volume of KV cache that must be maintained concurrently, further intensifying the memory burden. In addition, the growing prevalence of Mixture-of-Experts (MoE) architectures~\cite{MoE} has substantially increased the total number of model parameters. For example, from Mixtral 8$\times$7B~\cite{jiang2024mixtralexperts} (8 experts, 47B total parameters) to DeepSeek-V3~\cite{deepseekai2025deepseekv3technicalreport} (256 routed experts, 671B total parameters), the number of experts and the overall parameter count have grown substantially. Consequently, the confluence of increasing sequence lengths, large-scale batching, and MoE scaling has elevated memory capacity to a critical constraint in LLM inference, on par with memory bandwidth.

However, the pace of DRAM technology advancement has lagged considerably behind the improvements in computational throughput delivered by modern xPU (e.g., GPU, NPU)~\cite{ma2026challenges,10477550}. For instance, over the past 20 years, the peak compute capability of server-grade AI hardware has increased by 60,000$\times$ (scaling at 3.0$\times$ every 2 years). In stark contrast, DRAM bandwidth has only grown by 100$\times$ (scaling at 1.6$\times$ per two years), and single-chip memory capacity has merely scaled at a rate of 2$\times$ every two years~\cite{10477550}. While the memory bandwidth and capacity requirements of models and workloads are increasing exponentially, DRAM technology scaling is gradually saturating as it enters the 10nm-class nodes~\cite{10631320}. This vast and diverging disparity has exacerbated the so-called ``Memory Wall,'' where the limited capacity and bandwidth of data transfer, rather than compute capability, have become the primary bottleneck for AI applications, particularly in LLM inference~\cite{ma2026challenges,10477550}. To bridge this gap, various Near-Memory Processing (NMP) architectures have recently been proposed in both industry and academia to minimize data movement and perform computations close to the memory~\cite{sshbmpim1,sshbmpim2,aim1,aim2,aim3,neupims,attacc,duplex,h2llm,stratum,pimba}.

While hardware solutions like NMP strive to bridge the memory gap, the LLM architecture itself is rapidly evolving to inherently bypass these limitations algorithmically. Initially, to alleviate the memory burden caused by the KV cache within the attention mechanism, grouped-query attention (GQA) has been widely adopted~\cite{gqa}. However, GQA reduces the size of the KV cache but does not eliminate its fundamental growth with sequence length—leaving the quadratic complexity of attention and the proportional increase in KV cache. Pushing beyond these optimizations, hybrid architectures that replace a portion of the attention layers with state-space model (SSM) layers, such as Mamba-2~\cite{mamba2}, are rapidly gaining traction~\cite{waleffe2024,mamba_llama,ibm_granite_4_0_micro,bamba9b2025,nvidia2025nemotronhfamilyaccurateefficient,nvidia2025nvidianemotron3efficient,nvidia2025nemotron3nanoopen,samba,samba_v2,zamba2,hunyuan_turbos,falconh1,falconh1r}. SSMs maintain a fixed-size recurrent state regardless of the sequence length, thereby sidestepping the issue of KV cache growth that scales proportionally with sequence length. Additionally, the linear computational complexity of SSMs with respect to sequence length alleviates the quadratic overhead inherent in attention mechanisms, enabling significantly higher computational efficiency. Concurrently, MoE architectures—already a major driver of memory capacity demand as noted above—are trending toward finer granularity. By scaling up the number of experts while keeping the per-token activation count small, models can enhance expert specialization and overall model quality without proportionally increasing inference compute cost. As a result, there is a noticeable shift toward hybrid LLMs incorporating interleaved attention, SSM, and fine-grained MoE layers. For instance, NVIDIA's Nemotron3-Nano and Nemotron3-Super mix GQA layers, Mamba-2 layers, and MoE layers with 128 and 512 experts, respectively.~\cite{nvidia2025nvidianemotron3efficient,nvidia2025nemotron3nanoopen,nvidia_nemotron_3_super}.

\begin{figure}[t]
    \centering
    \includegraphics[width=0.99\columnwidth]{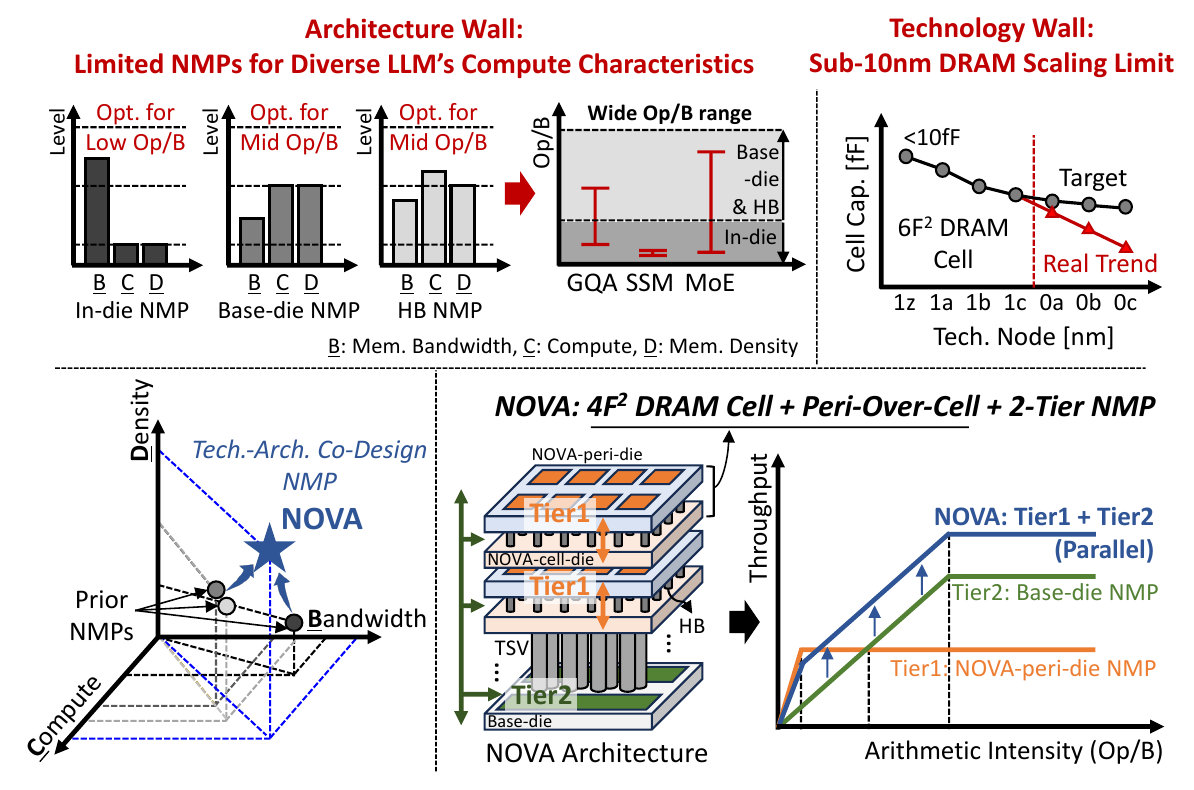}
    \caption{Overview of NOVA.}
    \label{fig1}
\end{figure}

However, the emergence of deeply interleaved hybrid LLM architectures and massively scaled fine-grained MoEs introduces two fundamental blockers for existing NMP architectures, as illustrated in Fig. 1. The first challenge is the \textbf{Architecture Wall}, which stems from the extremely heterogeneous compute characteristics. In hybrid LLMs, layers with vastly different memory access patterns and arithmetic intensities (Op/B) are tightly interleaved. For example, SSM is heavily memory-bound with very low Op/B, whereas GQA often exhibits higher Op/B. Fine-grained MoE further complicates execution due to highly dynamic and sparse access patterns, leading to wide fluctuations in Op/B. However, existing NMP solutions, such as in-die NMP (or processing-in-memory (PIM))~\cite{sshbmpim1,sshbmpim2,aim1,aim2,aim3,neupims,attacc,pimba}, base-die NMP~\cite{duplex}, and hybrid bonding (HB) NMP~\cite{h2llm,stratum}, are typically tailored to a specific Op/B range. As shown in Fig. 1, in-die NMPs are optimized for low Op/B, while base-die and HB NMPs are optimized for mid-range Op/B. Consequently, prior NMPs inevitably lead to severe hardware under-utilization and inefficiency when executing the diverse operations of hybrid LLMs.

The second challenge is the \textbf{Technology Wall}, driven by the physical scaling limits of DRAM technology. The shift toward large-scale fine-grained MoEs and extensive batching demands an unprecedented explosion in memory capacity. To fulfill this demand, DRAM technology must aggressively scale beyond the 10nm-class nodes (i.e., into sub-10nm nodes such as 0a, 0b, and 0c). However, the conventional $6F^2$ DRAM cell structure is rapidly hitting its physical limits~\cite{8976234,10185290,10631320,11075100}. As the physical dimensions shrink, maintaining the minimum required cell capacitance becomes exceedingly difficult, leading to degraded sensing margins and critical data retention issues. As shown in the real trend in Fig. 1, $6F^2$ cells fall short of the target capacitance required for sub-10nm scaling~\cite{4f2_isscc,10631320}, making it impossible to meet the capacity demands of LLMs without a fundamental breakthrough in DRAM technology.

In this paper, we propose \textbf{NOVA} (\textbf{N}ext-generation \textbf{O}ver-cell \textbf{V}ertical \textbf{A}rchitecture), a novel NMP architecture grounded in a technology-architecture co-design that simultaneously addresses the sub-10nm DRAM technology scaling limits and the heterogeneous computational demands of hybrid LLM inference. Specifically, NOVA is built upon two synergistic innovations:
\textbf{(1) $\boldsymbol{4F^2}$ vertical channel transistor (VCT) DRAM cell with peripheral (peri)-over-cell (POC) structure.} NOVA adopts a $4F^2$ DRAM cell that replaces the conventional $6F^2$ DRAM cell structure, enabling continued cell scaling into sub-10nm nodes while maintaining sufficient cell capacitance and directly increasing memory density by integrating more cells within the same die area. Furthermore, the POC structure relocates peripheral circuits to a peri-die stacked above the DRAM cell-die, reclaiming die area previously occupied by peripherals and further boosting memory density.
\textbf{(2) 2-tier NMP architecture with peri-die and base-die.} To address the Architecture Wall, NOVA introduces a 2-tier NMP design built upon the POC structure, which enables processing units (PUs) to be placed in the peri-die without any memory capacity reduction. NOVA vertically bonds a peri-die and a cell-die via HB to form a NOVA-die pair, and stacks multiple NOVA-die pairs on top of the shared base-die. Tier-1 (peri-die NMP) is optimized for low-to-mid Op/B operations such as SSM and cold-expert computation, while Tier-2 (Base-die NMP) targets mid-to-high Op/B operations, including GQA and hot-expert computation. By executing Tier-1 and Tier-2 in parallel, NOVA maximizes throughput across the full Op/B spectrum.
Through this co-design, NOVA efficiently accelerates the memory-bound decode phase operations of hybrid LLM inference—spanning Attention, SSM, and MoE—within a single NMP architecture, while simultaneously securing sufficient memory capacity to accommodate the massive KV cache and expert weights of fine-grained MoE models.
In summary, the key contributions of this paper are as follows:
\begin{itemize}
    \item We propose NOVA, a novel NMP architecture co-designed with $4F^2$ DRAM and POC structure, achieving ~2$\times$ memory density at iso-area and enabling sub-10nm DRAM scaling.
    \item We introduce a 2-tier NMP comprising a peri-die and a base-die NMP. It features three execution modes, including cross-tier parallel execution, to efficiently cover the full Op/B spectrum of hybrid LLMs (GQA, SSM, MoE) with negligible area overhead and no memory capacity loss.
    \item We evaluate NOVA across state-of-the-art hybrid and MoE LLMs. Compared to the GPU baseline, NOVA demonstrates significant performance gains, delivering an average of 4.5$\times$ higher throughput, 69.8\% lower end-to-end latency, and 5$\times$ better energy efficiency.
\end{itemize}

%% file: Outline/2_background.tex
\begin{figure}[t]
    \centering
    \includegraphics[width=0.99\columnwidth]{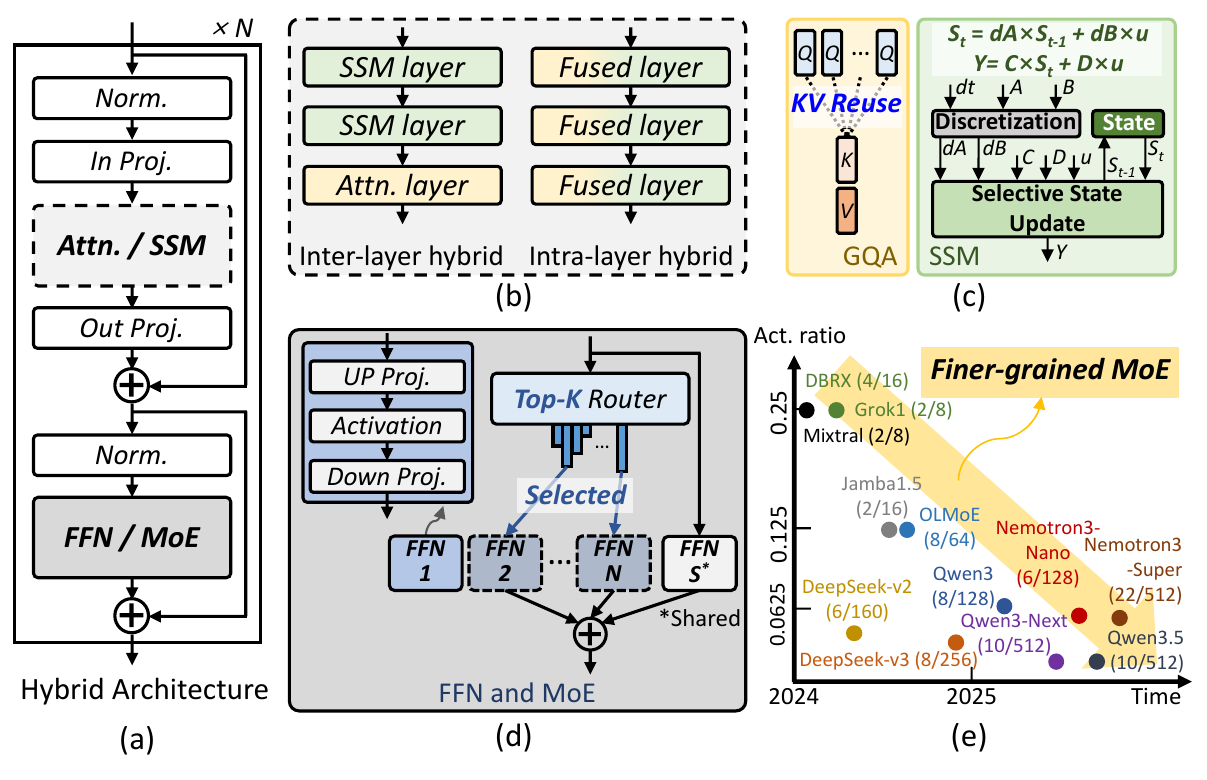}
    \caption{Architectural overview and operational description of modern hybrid LLMs. (a) General hybrid model structure. (b) Inter-layer vs. intra-layer hybrid designs. (c) GQA and SSM (Mamba-2) operations. (d) MoE layer with top-k routing. (e) Trend toward finer-grained MoE with decreasing activation ratios.}
    \label{fig2}
\end{figure}
\begin{figure}[t]
    \centering
    \includegraphics[width=0.99\columnwidth]{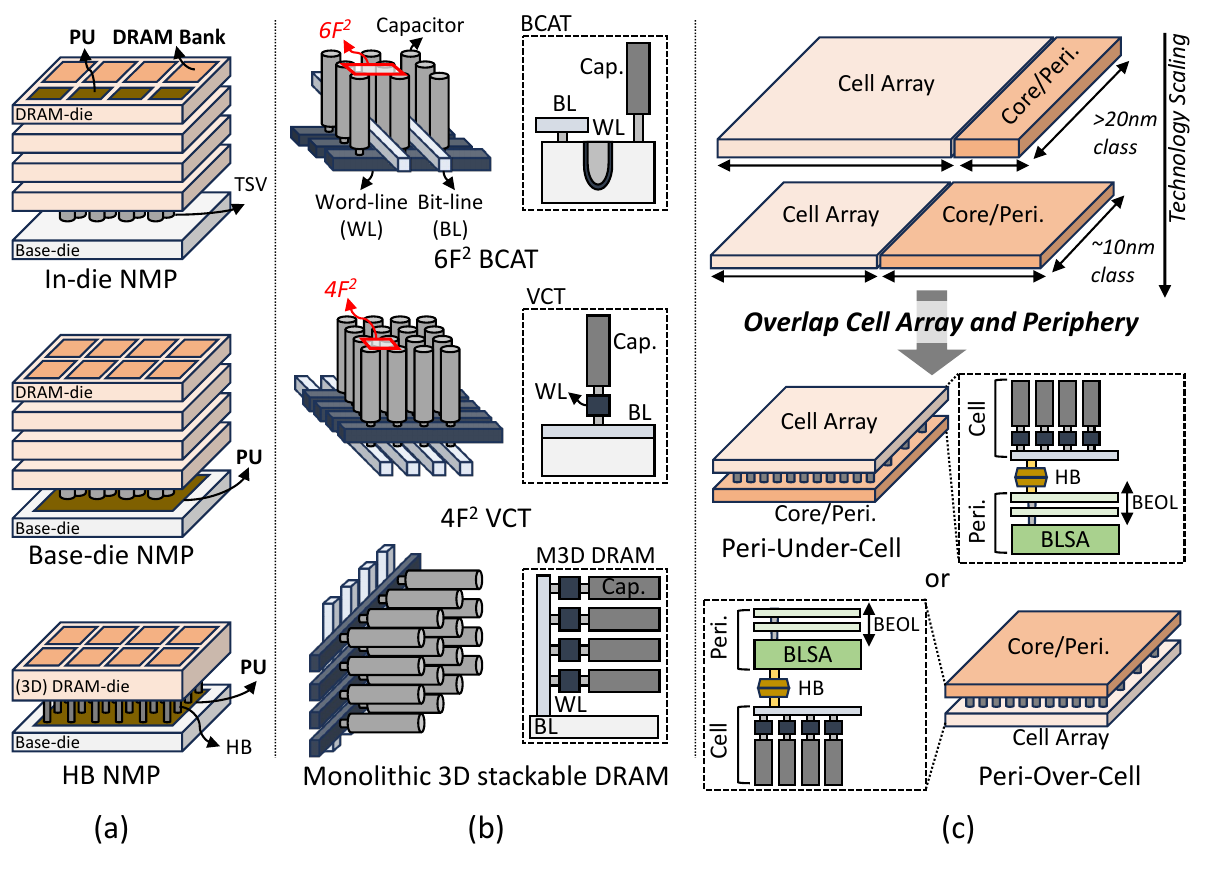}
    \caption{Prior NMP architectures and emerging DRAM technologies. (a) In-die, base-die, and HB NMP. (b) Evolution from conventional ${6F^2}$ BCAT to ${4F^2}$ VCT cell and monolithic 3D stackable DRAM cell for sub-10nm nodes. (c) Wafer-level vertical integration (Peri-Under-Cell or Peri-Over-Cell) via hybrid bonding, vertically overlapping the cell array and periphery.}
    \label{fig3}
\end{figure}

\section{Background}\label{Background}
\subsection{Heterogeneity in Modern LLMs}
The Transformer~\cite{vaswani2023attentionneed} has served as the dominant backbone of modern LLMs. However, its self-attention mechanism suffers from quadratic computational complexity with respect to sequence length and significant memory overhead due to the KV cache. To overcome the limitations of Transformers, SSMs such as Mamba~\cite{mamba1,mamba2} have recently emerged. By selectively updating input tokens based on their importance and compressing prior context into a fixed-dimensional hidden state, SSMs achieve linear computational complexity and constant memory footprint during inference. However, SSMs still lag behind Transformers in tasks such as in-context learning and exact recall~\cite{waleffe2024,arora2025simplelinearattentionlanguage} because their fixed-size state inevitably leads to gradual information decay over long sequences. These complementary characteristics of attention and SSM have driven the emergence of hybrid Transformer-Mamba architectures that mutually compensate for their respective limitations (see Figs. 2(a)-(c)), thereby achieving both high accuracy and computational efficiency~\cite{jamba,jamba1_5,waleffe2024,mamba_llama,ibm_granite_4_0_micro,bamba9b2025,nvidia2025nemotronhfamilyaccurateefficient,nvidia2025nvidianemotron3efficient,nvidia2025nemotron3nanoopen,samba,samba_v2,zamba2,hunyuan_turbos,falconh1,falconh1r,hymba,meta_hybrid}. 

Hybrid LLMs are generally classified into two categories: inter-layer and intra-layer~\cite{meta_hybrid}. Inter-layer hybrids, such as Nemotron3-Nano~\cite{nvidia2025nemotron3nanoopen}, interleave attention and SSM layers at configurable ratios. Conversely, intra-layer hybrids, like Hymba and Falcon-H1(R)~\cite{hymba,falconh1,falconh1r}, fuse attention and SSM heads within a single layer. Both approaches demonstrate that hybrid models can match or exceed the accuracy of homogeneous models while delivering substantially higher inference throughput~\cite{meta_hybrid}. In parallel, MoE has been increasingly integrated into hybrid architectures to scale model capacity without proportionally increasing inference costs~\cite{jamba1_5,nvidia2025nvidianemotron3efficient,nvidia2025nemotron3nanoopen,nvidia_nemotron_3_super}. At its core, an MoE layer replaces a standard dense feed-forward network (FFN) with a routing mechanism and a set of independent expert FFNs, as illustrated in Fig. 2(d). During execution, the router dynamically assigns each token to only the top-$k$ most relevant experts. This selective activation has enabled recent models to evolve toward finer-grained configurations (see Fig. 2(e)), activating a select few out of hundreds of available experts.

\subsection{Existing NMP Architectures}
NMP architectures have evolved into three main categories based on the placement of PUs and interconnect technologies, as shown in Fig. 3(a). Initially, in-die NMP~\cite{sshbmpim1,sshbmpim2,aim1,aim2,aim3,neupims,attacc,pimba} integrates PUs directly within the DRAM-die. However, implementing logic in DRAM processes entailed fundamental limitations, namely, area overhead and reduced memory capacity. To overcome this, base-die NMP~\cite{duplex} emerged, placing more PUs on the HBM base-die and securing high internal bandwidth via Through-Silicon Vias (TSVs). Building on this, HB NMP~\cite{h2llm,stratum} has been proposed to maximize interconnect density between the DRAM-die and the base-die, with finer pitch than TSVs. However, as we detail in Section 3.3, existing NMP architectures across all three categories are designed for specific computational patterns and fail to address the heterogeneous demands of hybrid LLMs.

\subsection{DRAM Technology Scaling and Its Limits}
DRAM technology has been continuously scaled to improve density and reduce the cost per bit, evolving 1T-1C cell structures like the buried channel array transistor (BCAT) from an $8F^2$ to a $6F^2$ layout~\cite{10631320,10185290,11075100,11400565,8976234}. Recently, rapid advances in AI technologies, including LLMs, have driven a surging demand for higher memory bandwidth and capacity, placing immense pressure on the pace of DRAM development. In response to this demand, products based on the sixth-generation 10nm-class technology node (D1c, approximately 11.3nm~\cite{11400565}) have entered mass production~\cite{hbm4_isscc}. However, as DRAM approaches its physical limits, historical gains in cost reduction and cell efficiency have slowed significantly. In particular, the BCAT-based $6F^2$ cells used in current 10nm-class nodes face fundamental process limitations. These constraints lead to increased cell resistance and decreased capacitance, which severely degrade sensing margins and data retention characteristics~\cite{8976234,10185290,4f2_isscc}.

\begin{figure}[t]
    \centering
    \includegraphics[width=0.99\columnwidth]{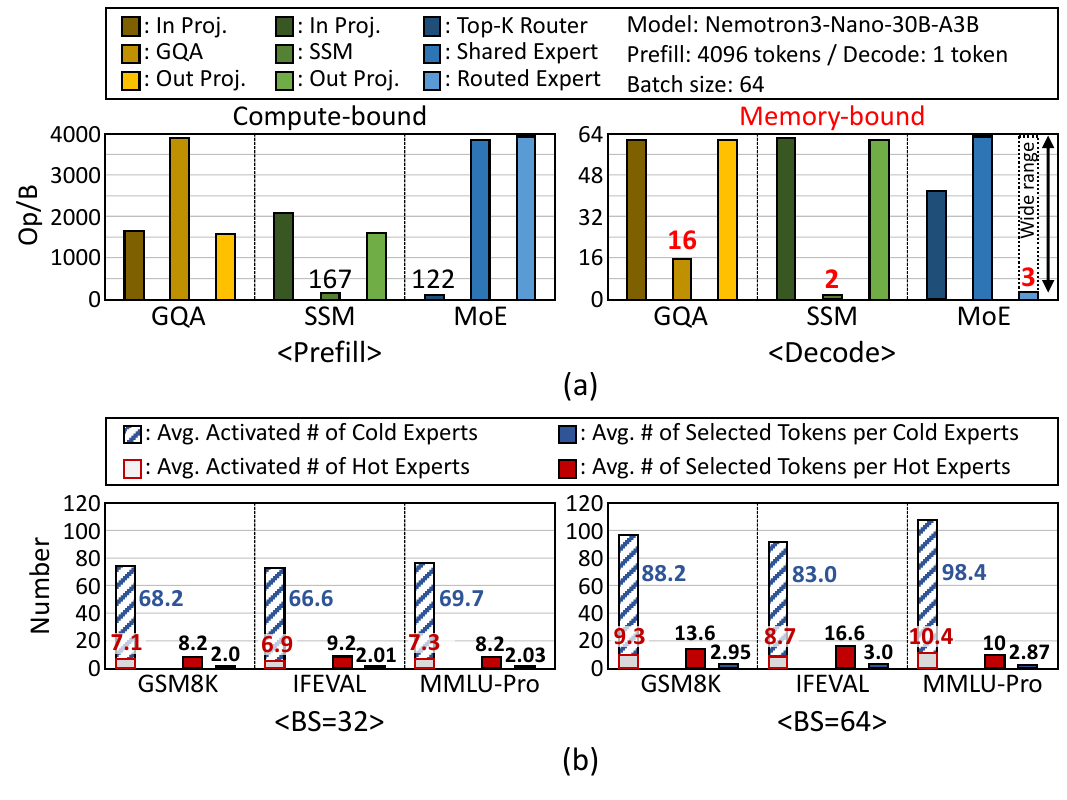}
    \caption{(a) The Op/B of GQA, SSM, and MoE layers of prefill and decode phases. (b) Average number of hot/cold experts and their per-expert token counts, evaluated on three benchmarks with batch sizes of 32 and 64.}
    \label{fig4}
\end{figure}

\subsection{Emerging DRAM Technologies}
To overcome the physical limitations of conventional $6F^2$ cell scaling and advance toward sub-10nm processes, two major categories of emerging technologies have been actively explored: cell-level structure innovations (Fig. 3(b)) and wafer-level structure innovations (Fig. 3(c)).
On the cell structure side, the $4F^2$ VCT cell has been proposed as a successor to the $6F^2$ BCAT cell~\cite{10145977,11075066,4f2_isscc,4f2_ss}. By orienting the channel vertically, the VCT decouples the access transistor footprint from the cell area, achieving a compact $4F^2$ layout while maintaining sufficient cell capacitance, mitigating short-channel effects, and preventing charge loss from row-hammer accesses due to its bulk-less structure~\cite{4f2_isscc}. Separately, monolithic 3D stackable DRAM has been explored as an alternative path to increasing density by vertically stacking multiple cell layers within a single die~\cite{10631471,10185290}, offering a potential route to higher capacity without relying solely on lateral scaling.
At the wafer level, advances in wafer bonding technologies have enabled structural separation between the cell array and peripheral circuits (peri-under-cell (PUC) or POC), both fabricated in DRAM processes. For example, \cite{10631320} fabricates the $6F^2$ cell array and peripheral circuits on distinct wafers and bonds them into a POC structure, so that the peripheral area fully overlaps the cell array. This eliminates the die area previously occupied by peripheral circuits, enabling significant chip size reduction without any design rule scaling. In addition, \cite{11075066} and \cite{4f2_isscc} combine the wafer-level separation with the $4F^2$ VCT cell in a PUC structure via wafer bonding or hybrid bonding, demonstrating that cell-level and wafer-level innovations can be jointly applied to maximize memory density.

%% file: Outline/3_motivation.tex
\begin{figure}[t]
    \centering
    \includegraphics[width=0.99\columnwidth]{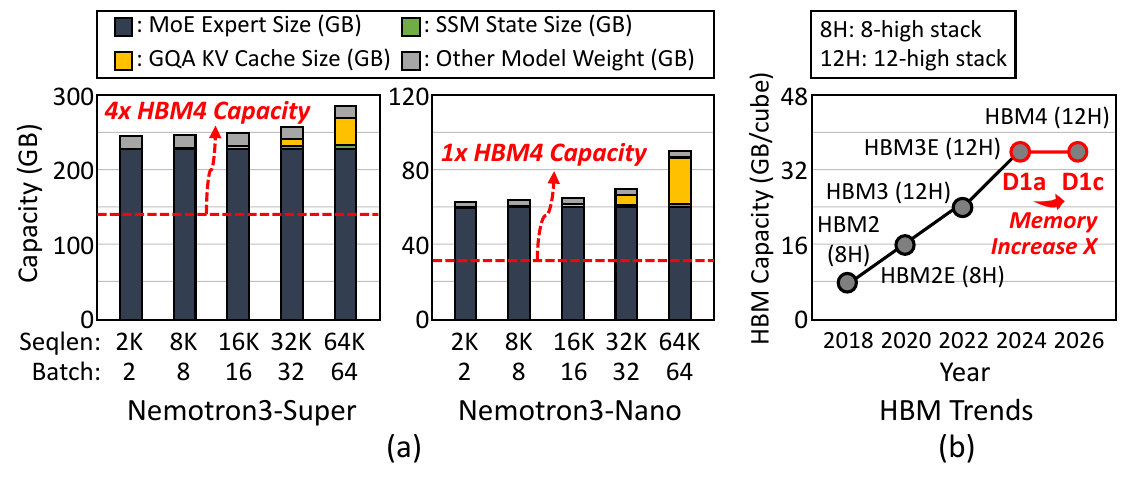}
    \caption{(a) Memory breakdown of Nemotron3-Super-120B-A12B (512 experts) and Nemotron3-Nano-30B-A3B (128 experts) under co-scaled sequence length and batch size. (b) HBM capacity scaling trends per cube~\cite{hbm4_isscc}.}
    \label{fig5}
\end{figure}

\section{Motivation}\label{Motivation}
\subsection{Heterogeneous Op/B in Hybrid LLMs}
Hybrid LLMs interleave GQA, SSM, and MoE layers to achieve both high accuracy and efficiency. However, these layer types exhibit fundamentally different computational characteristics. As shown in Fig. 4(a), during the decode phase of Nemotron3-Nano, Op/B values are uniformly low compared to the prefill phase, confirming that decode is severely memory-bound. Moreover, the Op/B varies significantly across layer types even within the same phase: GQA achieves an Op/B of approximately 16 owing to its 16:1 KV cache reuse ratio, whereas SSM exhibits an Op/B of only about 2 due to its recurrent state updates and frequent element-wise operations with additional state dimensions that inflate intermediate data sizes. 

MoE operations further exacerbate this heterogeneity in two ways. First, as discussed in Section 2.1, state-of-the-art MoE models are becoming increasingly fine-grained: the total number of experts grows while the activation ratio remains the same or decreases. This reduces the average number of tokens assigned to each expert at a given batch size, thereby lowering the average Op/B. For example, in Mixtral 8$\times$7B, which activates 2 out of 8 experts, Op/B increases with batch size and reaches approximately 20~\cite{duplex}, while the average Op/B of Nemotron3-Nano (6 out of 128 experts) is only about 3 despite increasing batch size. Second, dynamic token routing creates a large Op/B variance within a single MoE layer. Fig. 4(b) illustrates this imbalance: defining the top 10\% most-selected experts as hot, each hot expert processes approximately 5$\times$ more tokens than each cold expert, consistently across three benchmarks and two batch sizes. This skewed distribution means hot experts operate at a significantly higher Op/B than cold experts within the very same layer, further widening the computational heterogeneity.

\subsection{Exploding Memory Capacity Demands}
Each component of hybrid LLMs introduces a critical memory capacity pressure challenge. As shown in Fig. 5(a), the KV cache of GQA still scales proportionally with sequence length and batch size, and the state size of SSM still grows with batch size. Most critically, as MoE models trend toward a larger number of experts, the total expert weights dominate the overall memory footprint (accounting for over 85\% of total capacity in both Nemotron3-Super and Nemotron3-Nano), and this portion remains constant regardless of workload configuration. However, as shown in Fig. 5(b), HBM capacity scaling continues to lag behind. From HBM3E to HBM4, the HBM capacity per cube remains constant even though the DRAM technology scales down from D1a node to D1c node. Worse, physical limits below the sub-10nm node are expected to further aggravate the capacity bottleneck. Therefore, a technological breakthrough to achieve higher memory density is essential. Among the potential solutions discussed in Section 2.4, monolithic 3D stackable DRAM proposes a vertical scaling approach that could achieve high memory efficiency. However, realizing this potential requires at least 30 layers to match the memory density of $6F^2$ BCAT cells in the D1b process~\cite{science_dram}, whereas the most advanced prototypes demonstrated by DRAM vendors to date comprise only 5 layers~\cite{10631471}. As a result, $4F^2$ cells currently remain the primary candidate for sub-10nm nodes.

\begin{table}[t]
    \centering
    \caption{Comparison of Existing NMP Architectures}
    \includegraphics[width=0.999\columnwidth]{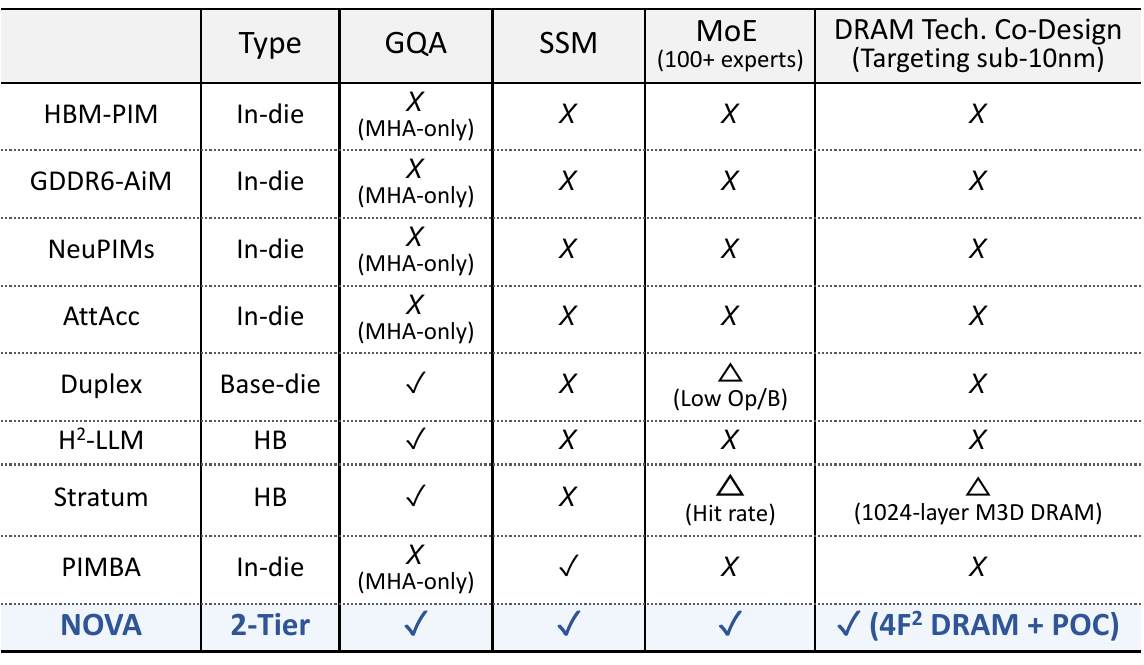}
    \label{tab:1}
\end{table}

\subsection{Limitations of Previous NMP Architectures}
As summarized in Table 1, none of the previously proposed NMP architectures addresses both challenges simultaneously.

\noindent
\textbf{In-die NMP.}
In-die NMPs from DRAM vendors (HBM-PIM~\cite{sshbmpim1,sshbmpim2}, GDDR6-AiM~\cite{aim1,aim2,aim3}) and academia (NeuPIMs~\cite{neupims}, AttAcc~\cite{attacc}, PIMBA~\cite{pimba}) integrate PUs near DRAM banks to accelerate memory-bound operations such as attention and SSM state updates. NeuPIMs and AttAcc offload multi-head attention (MHA) to in-die NMP to accelerate Transformers, while PIMBA offloads the selective state update within SSM layers to in-die NMP to accelerate post-Transformer models. However, in-die NMP suffers from several fundamental limitations. The restricted logic implementation capability of DRAM processes limits the compute-to-bandwidth ratio to only 1-2, rendering in-die NMPs fundamentally incapable of handling operations with higher Op/B. Moreover, the processing units occupy 20-27\% of the DRAM die area~\cite{duplex}, reducing memory capacity, which is a critical constraint given the exploding capacity demands of modern LLMs. 

\begin{figure*}[t]
    \centering
    \includegraphics[width=0.99\textwidth]{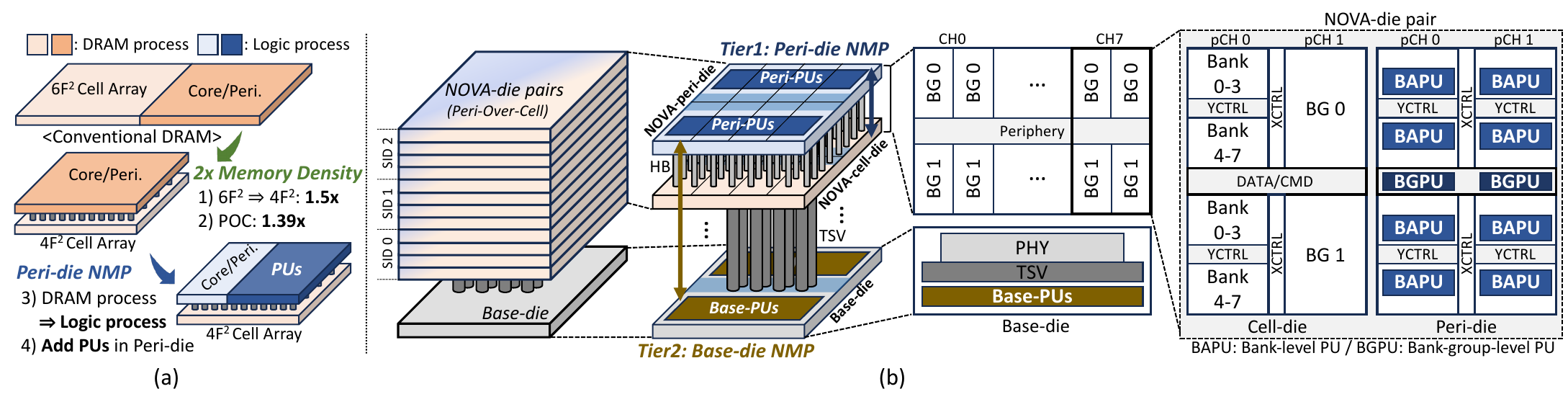}
    \caption {(a) Memory density improvement and logic-die NMP enablement through ${4F^2}$ VCT cell and POC structure. (b) Overall Architecture of NOVA.}
    \label{fig6}
\end{figure*}

\noindent
\textbf{Base-die NMP.}
Duplex~\cite{duplex} places more powerful PUs on the HBM base-die with additional TSVs to support GQA and MoE. However, it is constrained by the TSV pitch (22 $\mu$m) bandwidth limit, and its own energy-delay-area-product (EDAP) analysis indicates that in-die NMP is more efficient for Op/B $<$ 8, meaning it struggles to simultaneously cover low Op/B and medium-to-high Op/B workloads within a single architecture.

\noindent
\textbf{HB NMP.}
HB-NMPs leverage the fine pitch ($\sim$1\,$\mu$m) of HB to achieve significantly higher interconnect density compared to the base-die NMP. H\textsuperscript{2}-LLM~\cite{h2llm} achieves speedups through HB, however, it targets edge-level low-batch inference for standard Transformers without MoE layers. Stratum~\cite{stratum} also proposes an NMP architecture that connects a base-die to a monolithic 3D stackable DRAM via HB. By vertically stacking 1024 layers, Stratum achieves high memory density and optimizes MoE execution through in-memory tiering and expert replacement, accounting for the varying access latencies introduced by vertical stacking. However, monolithic 3D stackable DRAM has been demonstrated at only 5 layers~\cite{10631471}, and even state-of-the-art 3D NAND is capped at 300–400 layers~\cite{3DNAND_sk,3DNAND_samsung}. Furthermore, Stratum's speedup comes from keeping hot experts in a fixed-capacity fast tier. As MoE becomes fine-grained, token accesses spread across more experts, so the fast tier captures a smaller share of them, and the benefit shrinks. In addition, repeatedly reading and swapping experts within the same fast tier may raise endurance concerns, especially since 3D-stackable DRAM remains an emerging technology.

In summary, no existing NMP simultaneously overcomes both walls: even Duplex and Stratum degrade under fine-grained MoE for the mechanism-level reasons above, and no prior work exploits the architectural opportunities of emerging $4F^2$ VCT and POC structures—motivating our technology-architecture co-design.

%% file: Outline/4_NOVA_arch.tex
\section{NOVA: Technology-Architecture Co-Design}\label{NOVA_arch} 
To address the above challenges, we propose NOVA, which co-designs emerging DRAM technology and NMP architecture. On the technology side, NOVA achieves about 2$\times$ memory density at iso-area by replacing the cell structure and overlapping peripherals with the cell array. On the architecture side, NOVA repurposes the peri-die and leverages the base-die to introduce a 2-tier NMP that covers the full Op/B spectrum without any memory capacity loss.

\subsection{${4F^2}$ DRAM with POC Structure}
As illustrated in Fig. 6(a), NOVA is built upon two emerging DRAM technologies—the $4F^2$ VCT cell and the POC structure—that jointly achieve approximately 2$\times$ memory density over the conventional $6F^2$ BCAT cell while establishing the backbone for the proposed 2-tier NMP architecture. 
First, by reducing the cell area from $6F^2$ to $4F^2$, the memory density is increased by 1.5$\times$ based on an iso-area comparison. Second, whereas prior \cite{11075066} and \cite{4f2_isscc} adopt the PUC structure, NOVA adopts POC, stacking the core/peripheral circuits (e.g., the sub-word-line driver (SWD) and bit-line sense amplifier (BLSA))—which conventionally reside alongside the cell array—via HB so that the periphery fully overlaps the cell array. We choose POC over PUC because PUC must relay HB-transferred signals through the back-end-of-line (BEOL) of the peripheral wafer (see Fig. 3(c)), consuming about 30\% of its routing resources~\cite{4f2_isscc}, whereas POC avoids this overhead and thus offers higher area efficiency. Following prior POC~\cite{10631320} reports, this overlap reduces the chip area to about 72\% of the original, contributing an additional 1.39$\times$ density improvement; combined with the $4F^2$ cell, the overall density gain reaches 2.09$\times$ (1.5 $\times$ 1.39) at iso-area. The direct HB connection between each cell array (MAT) and the SWD/BLSA on the peripheral wafer further improves design flexibility and, as reported for such structures, reduces IDD0 and IDD5 by 10\% and 30\%, respectively, with no area overhead~\cite{4f2_isscc}.  

Building on this foundation, NOVA's key technology-side contribution is to repurpose the POC peri-die as an NMP substrate. Concretely, NOVA fabricates the peripheral wafer using a logic process rather than a conventional DRAM process. Even in the DRAM process-based POC structure, unused space already exists on the peripheral wafer~\cite{10631320}, but transitioning to a logic process amplifies this effect, as core/peripheral circuits can achieve equivalent performance in significantly less area, freeing up even more space on the peripheral wafer. NOVA populates this space with PUs to form the peri-die NMP. Since the PUs reside on the peri-die rather than the cell-die, NOVA introduces NMP capability without sacrificing any memory cell area, unlike in-die NMPs that forfeit over 20\% of the DRAM die area~\cite{duplex}. Also, the direct HB connection of all bit-lines to the peri-die's BLSA provides higher memory bandwidth and greater design flexibility.

\subsection{Overall Architecture of 2-Tier NMP}
Fig. 6(b) shows the 2-tier NMP architecture of NOVA, comprising a peri-die NMP (Tier-1) and a base-die NMP (Tier-2). NOVA consists of three types of dies: the cell-die, the peri-die, and the base-die. The cell-die and peri-die are vertically bonded via HB to form a single NOVA-die pair, and multiple NOVA-die pairs are stacked on the base-die. According to~\cite{4f2_isscc}, the HB integration of the $4F^2$ DRAM wafer and peripheral wafer increases the die thickness by only approximately 8\% compared to a conventional $6F^2$ DRAM-die. Moreover, unlike HBM4, which stacks DRAM dies using relatively coarse-pitch microbumps, NOVA employs fine-pitch HB for stacking and interconnection, keeping the total stack height comparable to that of HBM4. However, vertical stacking between NOVA-die pairs still requires TSVs for inter-die connectivity even when HB is used~\cite{TSV_ectc,TSV_vlsi}. Consequently, the memory bandwidth of Tier-2 is determined by the TSV interconnect rather than HB.

Each NOVA-die pair contains 8 channels (CH0–CH7), consistent with HBM4, and each CH is divided into two pseudo-channels (pCH). Within each pCH, two bank groups (BGs) are organized, each containing 8 banks. As shown in Fig. 6(b), two NMP tiers with independent bandwidth paths are overlaid on this hierarchy. Tier-1 accesses the cell-die directly beneath it via HB and integrates two granularities of PUs: a bank-level PU (BAPU) per bank and a bank-group-level PU (BGPU) per 2 bank-groups. BAPUs directly access the internal bandwidth of individual banks to perform bank-parallel memory-bound operations. BGPUs perform operations such as activation functions on intermediate results produced by the BAPUs and return them to the BAPUs for subsequent bank-level computation. This reduces unnecessary data movement to the base-die. Tier-2 consists of Base-PUs on the base-die, which access the cell-dies through TSVs. Starting from HBM4, the base-die is fabricated in an advanced logic process (4nm node)~\cite{hbm4_isscc}, and its ample area accommodates more powerful PUs than those on the peri-die. The difference in bandwidth-compute characteristics between the two tiers maps each to a distinct Op/B regime of hybrid LLMs. Tier-1 leverages the high internal bandwidth enabled by HB and is optimized for low-to-mid Op/B operations. Tier-2, while bandwidth-constrained by the TSV, provides substantial compute capability and is optimized for mid-to-high Op/B operations. Furthermore, since the two paths operate independently, Tier-1 and Tier-2 can process different operations concurrently, enabling parallel execution across tiers. This allows NOVA to efficiently cover the wide Op/B spectrum of hybrid LLMs within a single architecture. 

\begin{figure}[t]
    \centering
    \includegraphics[width=0.9\columnwidth]{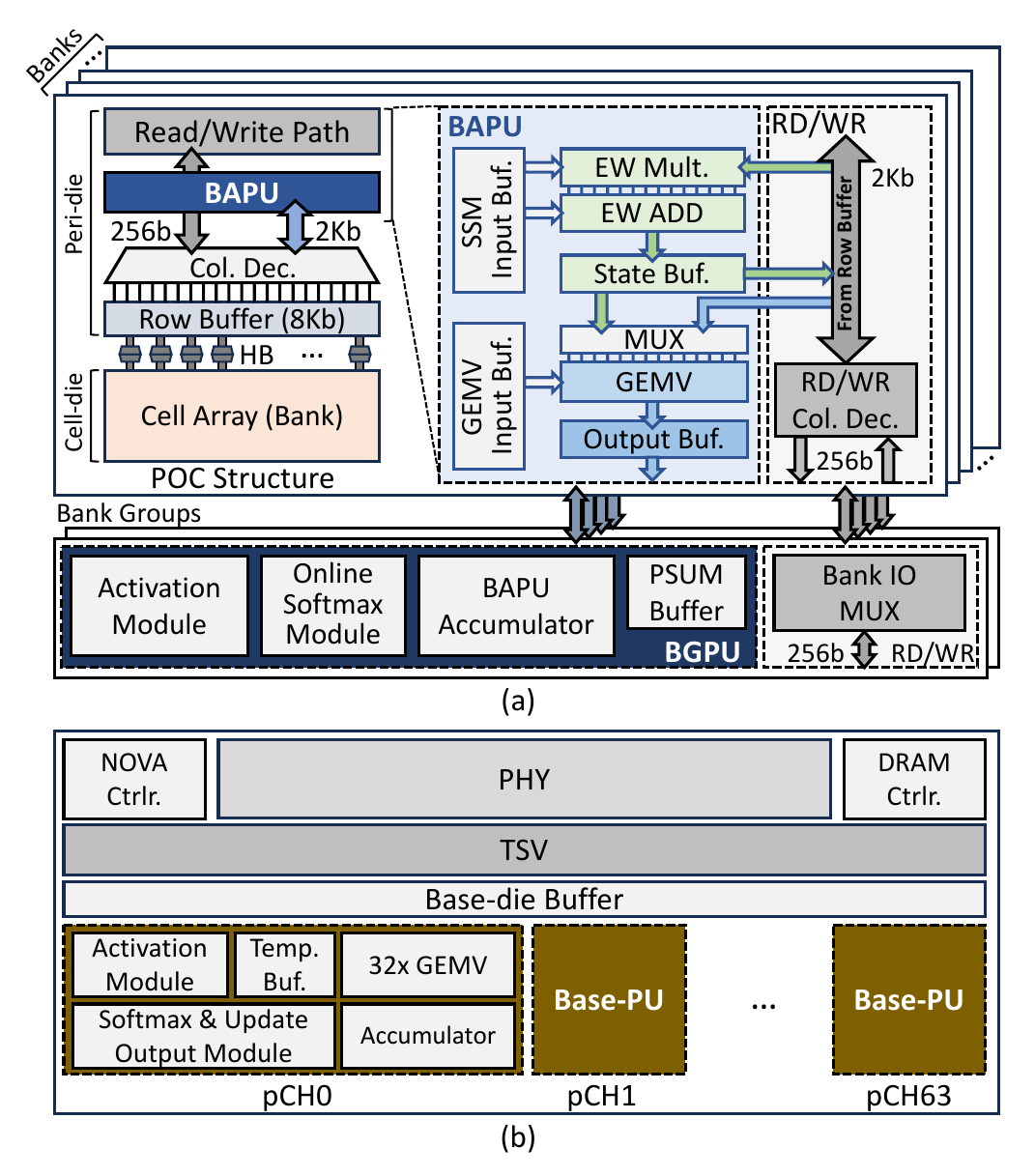}
    \caption {Microarchitectures of (a) peri-die NMP and (b) base-die NMP.}
    \label{fig7}
\end{figure}
\begin{figure}[t]
    \centering
    \includegraphics[width=0.99\columnwidth]{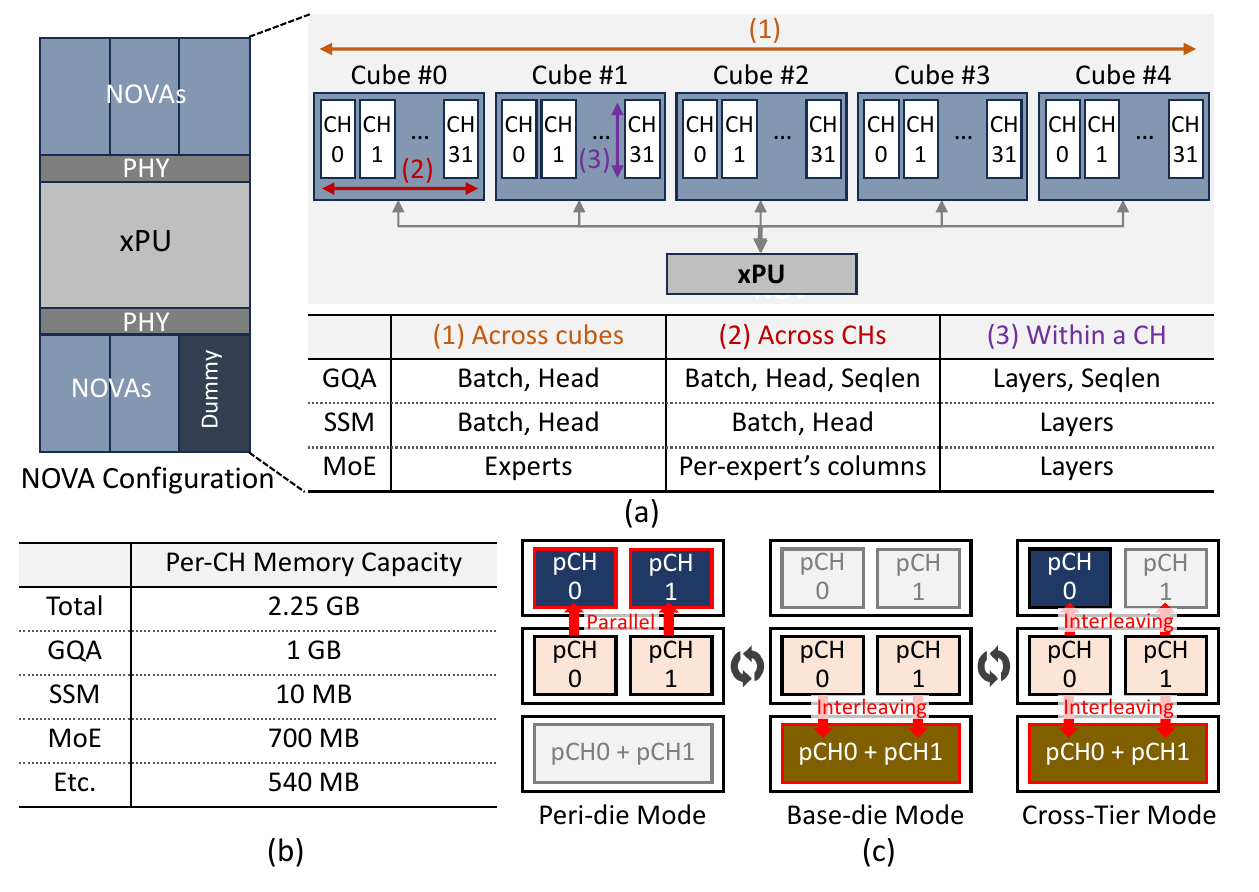}
    \caption {(a) Data mapping of GQA/SSM/MoE across cubes, channels, and within a channel, (b) per-CH memory allocation, and (c) three execution modes.}
    \label{fig8}
\end{figure}

\subsection{Tier-1 and Tier-2 Microarchitectures}
Figs. 7(a) and (b) present the microarchitectures of peri-die NMP and base-die NMP, respectively. Since the peri-die NMP is built upon the POC structure, data from an activated row in the cell-die is transferred through HB to the row buffer (8Kb) on the peri-die, where it is split into two paths: one to the BAPU for near-memory computation and another for the read/write interface. This dual-path access is realized through a two-level hierarchical column decoder structure: 2Kb of data is delivered to the BAPU to ensure high bank-level memory bandwidth while an additional column decoder provides 256b-granularity data transfer for the read/write path, maintaining compatibility with conventional HBM4. Furthermore, to provide higher memory bandwidth for the base-die NMP, NOVA bypasses the conventional BG IO mux between two 256b BG output data. Instead, it leverages BG-level parallelism and the increased TSV count from HB-TSV stacking to establish a 512b-wide data path, ensuring sufficient bandwidth for Tier-2 operations. This flexibility is enabled by the logic-process-based POC structure, as described in Section~4.1.

Each BAPU supports two types of operations. The first is the SSM state update, which includes an element-wise multiplier (EWM), an element-wise adder (EWA), an SSM input buffer, and a state buffer for storing updated states. Following the state update equations shown in Fig.~2(c), the EWM and EWA first update the state. Then, the updated state is simultaneously fed into a GEMV operation to compute the final SSM output and is written back to the cell array for the next time step's state update. This enables efficient state update operations with minimal data movement and bank-level parallelism, eliminating the need to transfer states to the base-die or xPU for computation and back to DRAM at every time step. The second is the GEMV, comprising a GEMV input buffer, GEMV unit, and output buffer. This path reads weight data from the row buffer at 2Kb-granularity to perform bank-parallel GEMV operations, serving not only SSM output computation but also GQA and MoE FFN operations. The BGPU processes the BAPU results from each bank to support softmax, activation, and accumulation operations required by GQA and MoE, and includes a BAPU accumulator, a partial-sum (PSUM) buffer, an online softmax module, and an activation module. By completing post-GEMV nonlinear operations entirely within the peri-die through the BGPU, unnecessary data movement to the base-die or xPU is minimized.

As shown in Fig. 7(b), the Base-PU resides on the base-die, accesses the NOVA-die pairs through TSVs, and is identically replicated across each pCH to support pCH-level parallel execution. Each pCH contains 32 GEMV units, providing significantly higher compute throughput than a BAPU. Each Base-PU is also equipped with a temporary buffer, an accumulator, an activation module, and a softmax and update output module. The temporary buffer stores data received through the TSVs from the two pCHs in a ping-pong fashion—receiving data from one pCH while the other is being accessed—and holds intermediate results, while the accumulator aggregates partial sums from the multiple GEMV units. The activation module handles nonlinear functions for MoE FFN layers, and the softmax and update output module computes the softmax over GQA scores and then updates the output. 

%% file: Outline/5_NOVA_mapping.tex
\section{NOVA's Operational Mapping and Execution}\label{NOVA_mapping} 
This section describes how GQA, SSM, and MoE operations are mapped and executed on the 2-tier NMP architecture presented in Section~4. We first present the data mapping and show that NOVA's memory capacity accommodates the entire decode working set, then define the three execution modes, and finally detail the per-operation execution flows.

\begin{figure}[t]
    \centering
    \includegraphics[width=0.9\columnwidth]{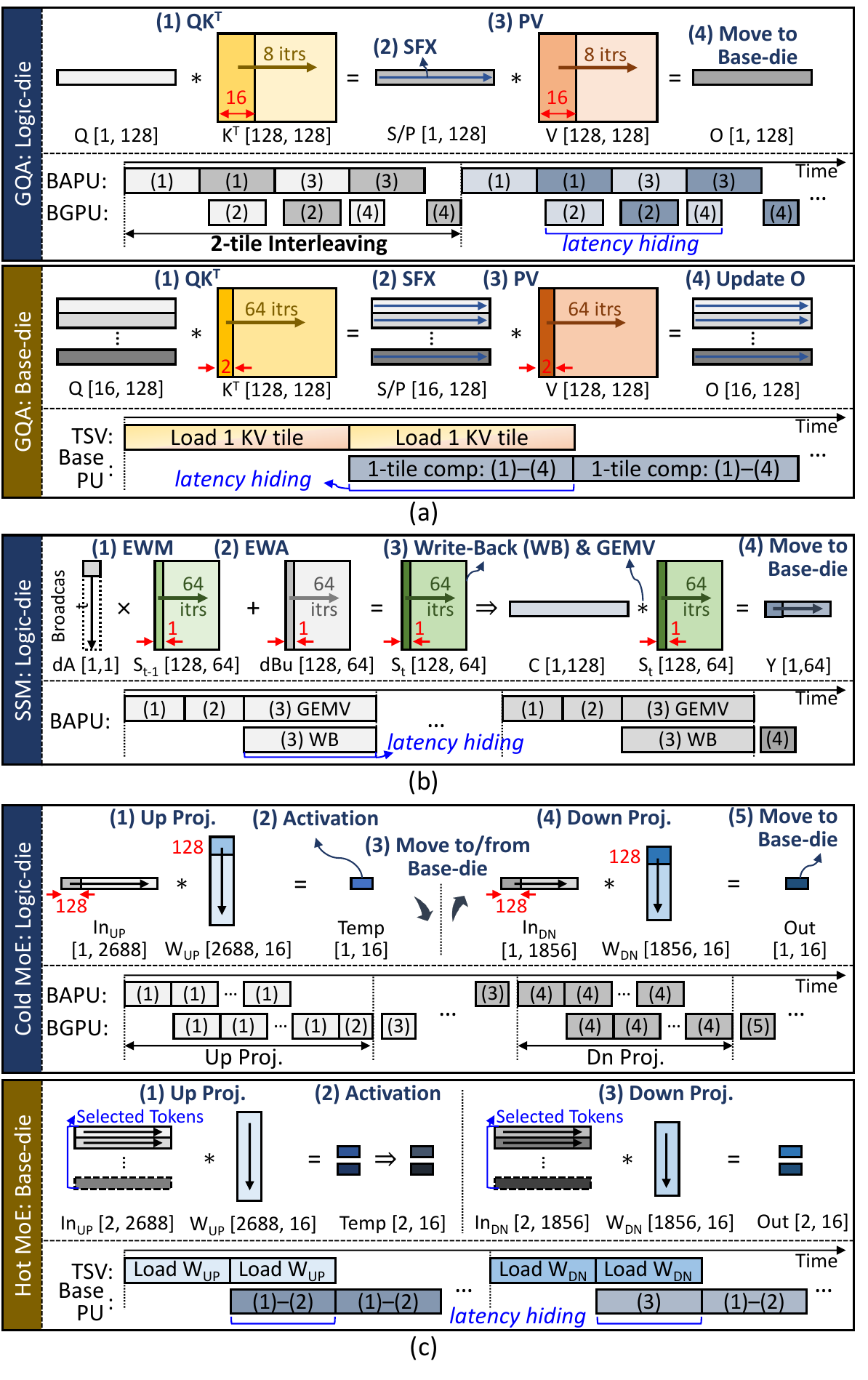}
    \caption {Per-pCH dataflow of (a) GQA, (b) SSM, and (c) MoE. This dataflow is based on the computational dimension of the Nemotron3-Nano-30B-A3B model.}
    \label{fig9}
\end{figure}

\subsection{Data Mapping and Capacity Provisioning}
Fig. 8(a) shows how NOVA distributes the data of each operation across its memory hierarchy. At the top level, data is partitioned across the five cubes along mutually independent dimensions—batch and head for GQA and SSM, and experts for MoE—so that NMP execution requires no inter-cube communication and per-cube outputs are simply gathered by the xPU. Since the CHs within a cube also operate in parallel, GQA and SSM are further partitioned across CHs by batch and head in the same manner, while the weights of the experts assigned to each cube are split column-wise across its CHs. Within a CH, the remaining dimensions (layers and sequence) are distributed over bank groups and banks so that all banks operate concurrently, maximizing bank-level parallelism. Moreover, this mapping is deterministic: locating the banks that hold a KV partition, a state slice, or a routed expert reduces to simple address arithmetic in the NOVA controller. Since both tiers access the same cell-dies, KV caches, states, and expert weights never need to be relocated at runtime; the only decision left at execution time is which tier computes on the data.
Fig. 8(b) shows how much GQA, SSM, and MoE data each CH accommodates. Owing to NOVA's 2$\times$ memory density, each CH provides 2.25 GB, of which 1 GB is allocated to the KV cache, 10 MB to SSM states, 700 MB to expert weights, and 540 MB to the remaining weights and activations. Therefore, the entire decode phase working set resides in NMP.

\subsection{Three Execution Modes}
Fig. 8(c) describes the three execution modes supported by NOVA. Depending on the Op/B characteristics of each operation, the appropriate mode is selected among Tier-1, Tier-2, and parallel cross-tier NMP. First, operations with very low Op/B, such as GQA with low KV reuse and SSM, are executed using only the peri-die NMP, fully exploiting bank-level parallelism through all-bank activation across all pCHs. Second, operations with sufficiently high Op/B, such as GQA with high KV reuse, are executed using only the base-die NMP, leveraging the high compute throughput of the Base-PU. Third, operations with moderate Op/B, such as GQA with moderate KV reuse and hot/cold MoE, utilize both tiers in parallel. Analogous to the pCH interleaving technique in conventional HBM, NOVA interleaves Tier-1 operations based on bank-parallel computation with Tier-2 operations that exploit BG-level parallelism. Specifically, while Tier-1 performs computation via all-bank activation, Tier-2 concurrently accesses two BGs in parallel via per-bank activation to transfer 512b-wide data to the base-die. By temporally interleaving the operations of both tiers, NOVA achieves high-throughput parallel execution of operations such as GQA and MoE.

\subsection{GQA/SSM/MoE Execution}
\noindent
\textbf{GQA Execution.}
Fig. 9(a) shows how GQA is executed across NOVA's two tiers. Since the Op/B of GQA varies depending on the KV reuse ratio, it is supported on both the peri-die and the base-die. To minimize softmax overhead and data movement while efficiently supporting GQA, the fused tiling technique from FlashAttention-2~\cite{fa2} is adopted. On the peri-die, attention operations (QK\textsuperscript{T}, softmax, PV) are performed per tile ([1, 128]) using the BAPU and BGPU. A proposed 2-tile interleaving technique overlaps the softmax computation time with GEMV operations such as QK\textsuperscript{T} and PV, further reducing latency. The completed tiles are sequentially transferred to the base-die, where the FlashAttention-2 update output operation corrects the per-tile attention results. The base-die runs the same tiled attention with higher throughput, overlapping KV-tile loads over TSVs with computation.


\noindent
\textbf{SSM Execution.}
Unlike GQA, SSM is an extremely low-Op/B operation and is therefore executed exclusively on the peri-die, as shown in Fig. 9(b). Input data required for the state update, such as $dA$ and $dBu$, are received from the base-die and preloaded into the SSM input buffer within the BAPU. The previous time step's state ($S_{t-1}$) is then read from the row buffer and updated through the EWM and EWA units. The updated state ($S_t$) is written back to the cell array for the next time-step state update operation and simultaneously forwarded to the GEMV unit to compute the SSM output with $C$ in parallel. This overlaps the write-back latency with the GEMV computation time, enabling efficient SSM execution. 

\noindent
\textbf{MoE Execution.}
Fig. 9(c) shows how MoE operations are executed across the two tiers. As shown in Fig. 4(b), the dynamic top-$k$ routing in MoE leads to a highly skewed distribution of tokens across experts, resulting in hot experts having significantly higher Op/B than cold experts. Accordingly, the peri-die handles cold-expert computation while the base-die handles hot-expert computation. For cold-expert computation, since the number of tokens assigned to each expert is very small, inputs are processed one at a time. The up projection is performed by sweeping GEMV operations in units of [1, 128] along the weight dimension, and the partial sums from each BAPU are accumulated using the BAPU accumulator and PSUM buffer in BGPU. After applying an activation function such as ReLU\textsuperscript{2}, the results are transferred to the base-die to aggregate the outputs distributed across pCHs. Once the up projection is complete, the input for the down projection is loaded into the GEMV input buffer within the BAPU, and the down projection GEMV operation is performed in the same manner. In contrast, hot experts consist of relatively few experts, each assigned a large number of tokens, enabling frequent weight reuse. This makes the base-die NMP, with its higher compute throughput, well-suited for hot-expert computation. Similar to GQA, the latency of loading up/down projection weights through TSVs is overlapped with the Base-PU computation time, enabling efficient hot-expert execution. 

\noindent
\textbf{Cross-Tier Traffic.}
The dataflow shown in Fig. 9 minimizes data movement between tiers. For SSM, since computations are exclusively performed in Tier-1, cross-tier traffic is strictly limited to exporting activations (such as $dA$, $dBu$, and $C$) and the final output $Y$ to the xPU. Similarly, in GQA, only the input $Q$ and output $O$ are transferred via TSVs. For MoE operations, inter-tier data movement through TSVs consists solely of input/output vectors and the intermediate data generated between the cold experts' projection operations. Given the 64 pCHs and the massive TSV bandwidth of 4 TB/s, the data payload per-pCH operation (512 B for GQA, 16 KB for SSM, and 17.75 KB for MoE) is negligible.


\begin{table}[t]
    \centering
    \caption{Model Configuration for Evaluation}
    \includegraphics[width=0.99\columnwidth]{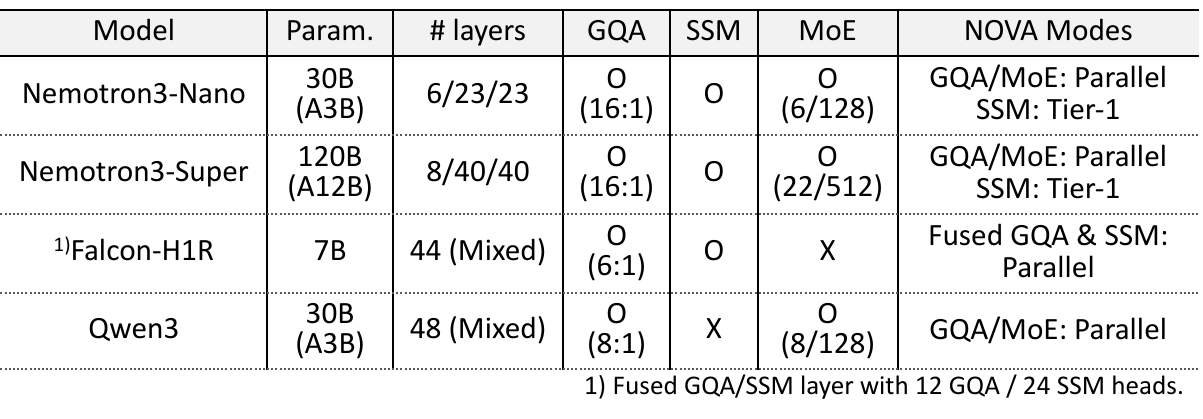}
    \label{tab:2}
\end{table}
\begin{table}[t]
    \centering
    \caption{HBM4 Configuration}
    \includegraphics[width=0.99\columnwidth]{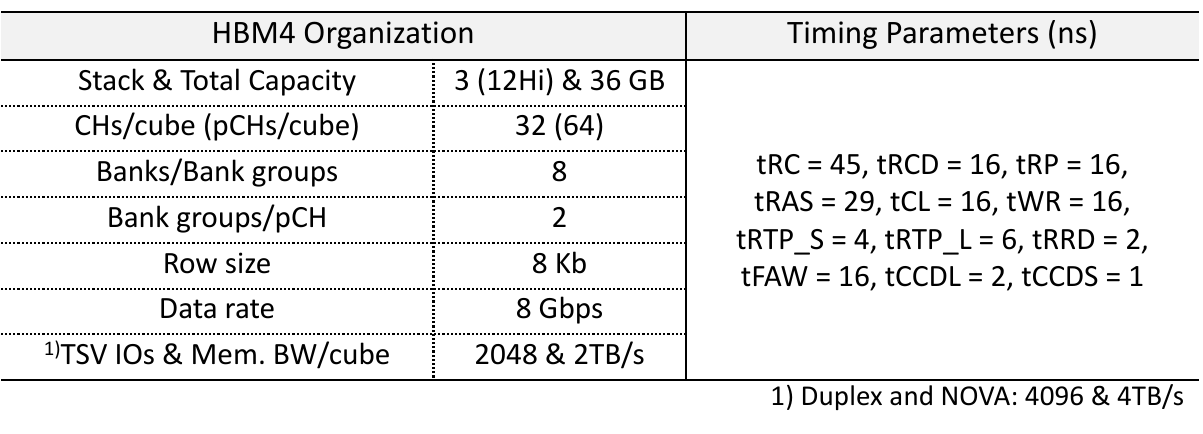}
    \label{tab:3}
\end{table}

%% file: Outline/6_evaluation.tex
\section{Evaluation}\label{eval} 
\subsection{Methodology}
\noindent
\textbf{Model.}
To evaluate NOVA, we use Nemotron3-Nano-30B-A3B~\cite{nvidia2025nemotron3nanoopen}, Nemotron3-Super-120B-A12B~\cite{nvidia_nemotron_3_super}, Falcon-H1R-7B~\cite{falconh1r}, and Qwen3-30B-A3B~\cite{yang2025qwen3technicalreport} (see Table 2). Nemotron3-Nano and Nemotron3-Super are representative hybrid MoE LLMs that incorporate GQA, SSM, and MoE layers, with 16:1 GQA and 128 and 512 experts, respectively. Falcon-H1R-7B is a representative intra-layer hybrid LLM in which GQA and SSM operations are mixed within a single layer using 6:1 GQA. This model is included to validate the effectiveness of parallel GQA/SSM execution enabled by the proposed 2-tier NMP. Finally, to demonstrate that NOVA also achieves efficient inference on pure Transformer-based MoE LLMs beyond hybrid architectures, we include Qwen3-30B-A3B, which employs 8:1 GQA with 128 experts. For each model, operations are mapped to the default tile granularity, accounting for model-specific configurations such as the hidden dimension, number of layers, and number of heads. Lastly, all models are evaluated using BF16 precision.

\begin{figure}[t]
    \centering
    \includegraphics[width=1\columnwidth]{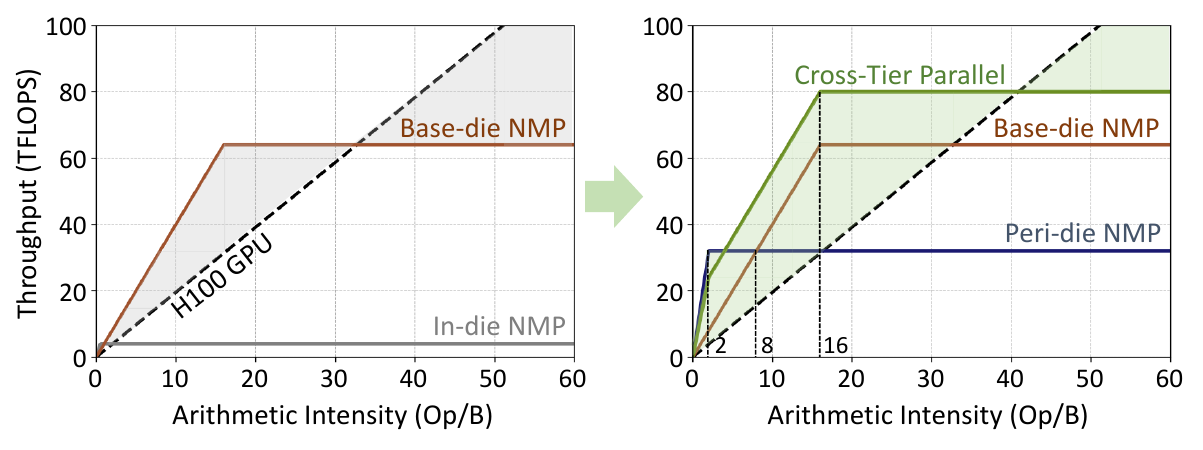}
    \caption {Roofline model of NOVA.}
    \label{fig10}
\end{figure}

\begin{figure*}[t]
    \centering
    \includegraphics[width=0.99\textwidth]{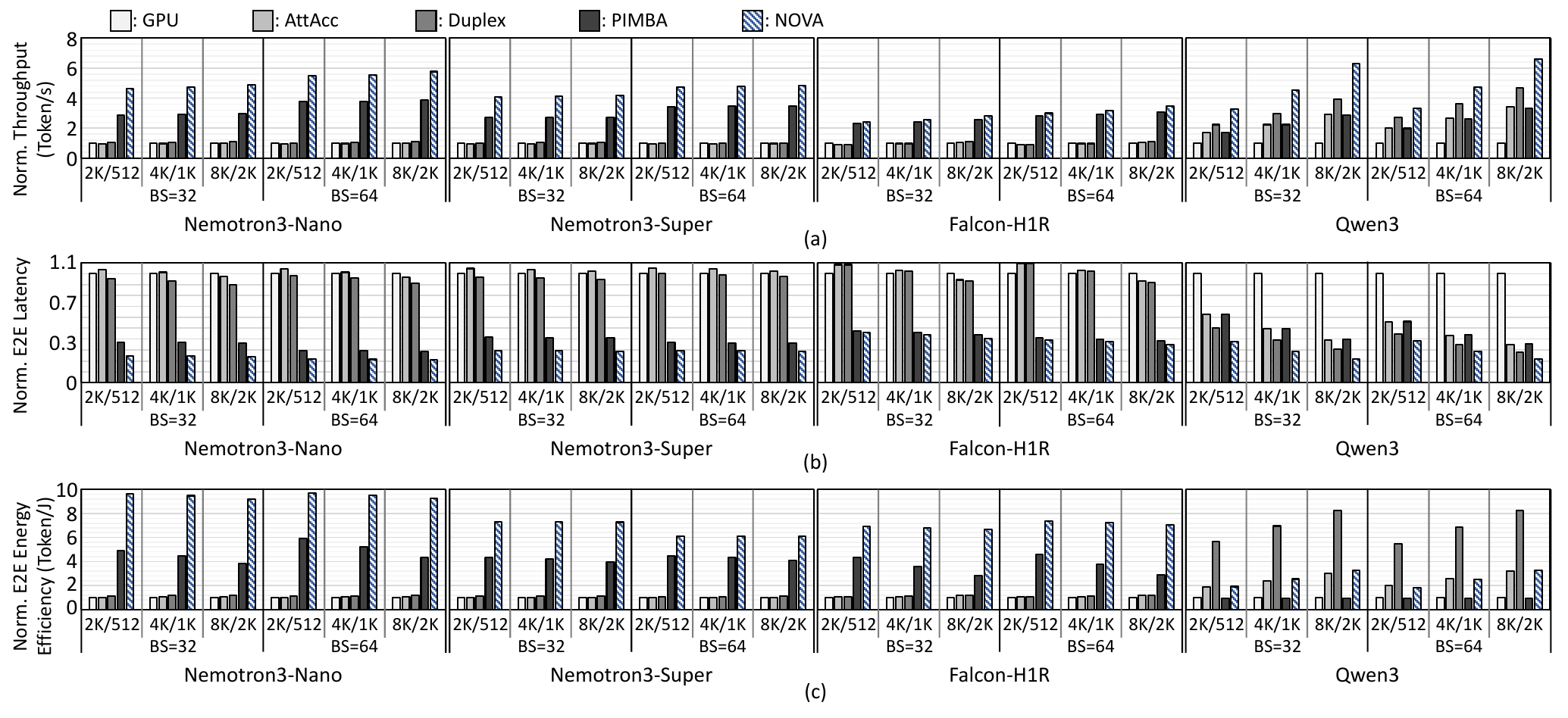}
    \caption {Evaluation and comparison of (a) system decoding throughput, (b) end-to-end latency, and (c) energy efficiency.}
    \label{fig11}
\end{figure*}

\noindent
\textbf{Baselines and Performance Evaluation.}
We compare NOVA against (1) H100 GPU, (2) AttAcc~\cite{attacc}, (3) Duplex~\cite{duplex}, and (4) PIMBA~\cite{pimba}. Only decode-phase GQA/SSM/MoE are offloaded to each NMP; all remaining operations, including prefill, run on the GPU. Operations unsupported by a baseline fall back to the GPU (AttAcc: SSM and MoE; Duplex: SSM; PIMBA: MoE), whereas NOVA executes all three on the NMP.
 
For a fair performance comparison, all NMP architectures, including NOVA, are implemented using an in-house cycle-accurate simulator built on Ramulator2~\cite{luo2023ramulator2}, following the AttAcc methodology~\cite {attacc}. 
Every configuration uses an NVIDIA H100 GPU paired with five HBM cubes, following the memory organization of the H100 (5$\times$16~GB), with each cube scaled up to HBM4 with 36~GB capacity for a total of 180~GB (for NOVA, a 2$\times$ per-cube density). Each cube is directly attached to the GPU through an independent channel with no cube-to-cube links (see Fig. 8(a)). Workloads are mapped cube-locally so that each GEMV completes within a single cube and only small activation results are gathered across cubes. The cost of this inter-cube aggregation and NMP$\leftrightarrow$GPU synchronization is modeled in our simulator and applied identically to all baselines.
The HBM is standardized to HBM4 specifications~\cite{hbm4_isscc, JEDEC_JESD270-4} (see Table 3) with 12-high stacking, 36 GB capacity, and a per-pin I/O data rate of 8 Gbps across all baselines. The timing parameters follow prior works~\cite{pimba,rome, folded_bank,hbm4_isscc,JEDEC_JESD270-4}.
NOVA and Duplex, both employing base-die NMP, use 4096 TSVs under iso-TSV conditions following Duplex~\cite{duplex}, while GPU and in-die NMP baselines use the conventional HBM4 configuration with 2048 TSVs. For PUs, AttAcc and PIMBA are configured with 16 BF16 GEMV units per bank to match the 256b per-bank read granularity~\cite{attacc}, preserving their original architectures (PIMBA additionally includes EWM and EWA for state update operations). The peri-die of NOVA is equipped with 128 BF16 GEMV units, EWM, and EWA per bank to support 2Kb-granularity operations, while the base-die incorporates 32$\times$ 128 BF16 GEMV units per pCH. Therefore, the peri-die and base-die NMP achieve the peak FLOPS of 32 TFLOPS and 64 TFLOPS for 2 Op/B and 16 Op/B, respectively. The base-die of Duplex is configured with the same memory bandwidth and compute throughput as that of NOVA. 

Experiments were conducted at batch sizes of 32 and 64, sweeping the input/output sequence lengths ($L_{in}$/$L_{out}$) across (2K/512), (4K/1K), and (8K/2K), and comparing throughput, latency, and energy efficiency. Also, considering the varying Op/B, we applied adaptive NOVA modes. Low Op/B operations like SSM are mapped exclusively to Tier-1, while GQA and MoE exploit parallel execution to maximize throughput. Lastly, for Nemotron3-Super, all evaluations are performed on a dual-GPU system interconnected via NVLink4 (900GB/s) due to the large model size.

\noindent
\textbf{Area and Energy Consumption.}
To estimate the area and energy consumption of NOVA, the PUs and SRAMs were designed at the RTL level in System Verilog and synthesized using Synopsys Design Compiler~\cite{synopsys_dc} at 14 nm technology, 0.8 V, and 1 GHz (matching the tCCDS of 1 ns). Energy consumption was obtained through post-synthesis simulation. We assume 7 nm and 4 nm processes for the peri-die and base-die, respectively, and scale the PU area and energy consumption accordingly~\cite{sarangi_iscas2021}.
For DRAM energy modeling, the POC structure offers several advantages over conventional HBM, including reduced wire length, lower wire RC delay, and lower supply voltage due to the transition from DRAM to a logic process for peripheral circuits. Reflecting these factors, we derive the energy for activation, read/write operations, and data movement based on prior work~\cite{fine_grained_dram,hbm4_isscc,4f2_isscc,folded_bank}.

\subsection{Roofline Analysis: 2-Tier Roofline}
To clarify how NOVA scales with Op/B and how it differs from a conventional NMP+xPU system, we analyze its roofline model in Fig. 10. Conventional in-die NMP and base-die NMP are each specialized for a specific Op/B region, so operations outside that region must be offloaded to the xPU. In contrast, NOVA forms two independent NMP rooflines. While retaining the base-die NMP (64 TFLOPS), which offers high compute throughput but relatively limited memory bandwidth, NOVA adds the peri-die NMP (32 TFLOPS), which secures higher compute capability than conventional in-die NMP and is thus more efficient even in the low-Op/B region. More importantly, because the two tiers use independent bandwidth paths (HB and TSV), cross-tier parallel execution can superimpose the two rooflines. Here, since the peri-die also uses part of its resources to transfer data to the base-die, the aggregated roofline is the base-die NMP combined with a portion of the peri-die NMP, achieving higher throughput than either tier alone (green region). As a result, NOVA covers a much wider Op/B range within the NMP itself, minimizing offloading while achieving higher performance—this is how NOVA overcomes the Architecture Wall.

\subsection{Performance Improvement}
\noindent
\textbf{Throughput.}
NOVA achieves significantly higher throughput compared to both the GPU baseline and state-of-the-art NMP architectures, as shown in Fig. 11(a). For the three hybrid models, NOVA delivers an average speedup of 4.2$\times$ (up to 5.8$\times$) over the GPU baseline, and 1.4$\times$ (up to 1.7$\times$) over PIMBA. In Qwen3, NOVA achieves an average improvement of 4.8$\times$ (up to 6.6$\times$) over the GPU and 1.43$\times$ (up to 1.6$\times$) over Duplex. This superior performance is primarily attributed to NOVA's ability to completely compute all three key operations (GQA, SSM, and MoE) on the NMP, thereby eliminating data movement overhead with the GPU. Furthermore, its cross-tier parallel execution significantly amplifies these performance gains. We also observe that NOVA attains even higher throughput as the sequence length ($L_{in}$/$L_{out}$) and batch size increase. As the $L_{in}$/$L_{out}$ grows, the KV cache size expands; similarly, larger batch sizes increase the number of states and the number of activated MoE experts. Consequently, the proportion of operations successfully processed within the NMP increases, thereby improving efficiency. Specifically, executing SSM on the GPU suffers from extremely low utilization and prolonged execution time. This is due to the massive intermediate data generated from tensor dimension expansion, intensive element-wise operations, and frequent DRAM accesses required for state updates. By effectively supporting SSM, PIMBA achieves significant performance gains over other baselines in hybrid models. Conversely, in Qwen3, which lacks SSM layers, Duplex—accelerating both GQA and MoE—shows relatively high performance. Notably, because Qwen3 has a high proportion of GQA operations, NOVA's improvement over the GPU becomes even more pronounced as $L_{in}/L_{out}$ scales up.

\noindent
\textbf{End-to-End Latency.}
As shown in Fig. 11(b), NOVA achieves the most significant reduction in end-to-end (E2E) latency. In the hybrid models, NOVA reduces latency by an average of 69.2\% compared to the GPU baseline, whereas AttAcc and Duplex show marginal improvements. Although these two partial offloading architectures efficiently accelerate GQA and GQA/MoE operations, respectively, the substantial SSM computations that remain on the GPU pose a severe bottleneck. Furthermore, compared with executing purely on the GPU, transitioning between the NMP and the GPU introduces communication and synchronization overhead at the start and end of offloaded operations. Fortunately, as $L_{in}$/$L_{out}$ and batch size increase, the NMP acceleration effect becomes more pronounced, thereby widening the latency reduction gap over the GPU baseline. In the case of Qwen3, while all baselines achieve some latency reduction due to the absence of SSM layers, NOVA still demonstrates the most superior efficiency (an average reduction of 70.3\%). Ultimately, NOVA provides stable, dramatic acceleration across model architectures and further reduces latency through the latency-hiding effect of its cross-tier parallel execution.

\noindent
\textbf{End-to-End Energy Efficiency.}
As shown in Fig. 11(c), the normalized E2E energy efficiency results exhibit a slightly different trend compared to the previously discussed throughput and E2E latency. Overall, by fundamentally eliminating the massive data movement energy between the HBM and the GPU, NOVA still achieves exceptional energy efficiency, delivering an average improvement of 7.5$\times$ over the GPU baseline. However, in the hybrid models, as $L_{in}$/$L_{out}$ and batch size increase, NOVA’s energy efficiency shows a slight downward trend, contrasting with its performance scaling. This is intrinsically due to the frequent state update operations of SSM. As the workload scales, the frequent DRAM internal activations required to update massive SSM states consume substantial dynamic power, partially offsetting the data movement energy savings. Notably, PIMBA suffers a more rapid drop in energy efficiency under heavier workloads. Because PIMBA operates at a smaller 256b-granularity state update compared to NOVA (2Kb), it generates an excessive number of DRAM activation commands. Conversely, in the Qwen3 model, Duplex achieves the highest energy efficiency with an average improvement of 6.7$\times$ over the GPU baseline, followed by NOVA with an average improvement of 2.4$\times$. This outcome stems from NOVA actively engaging its cross-tier parallel execution mode to maximize throughput and minimize latency, which naturally incurs higher active power consumption. Specifically, the peri-die, which performs highly parallel bank-level operations, activates eight times more banks compared to the base-die. In essence, NOVA demonstrates a deliberate performance-energy trade-off, exchanging a marginal amount of energy efficiency for overwhelming real-time performance, while still maintaining highly competitive energy savings compared to the GPU  baseline.

\begin{figure}[t]
    \centering
    \includegraphics[width=1\columnwidth]{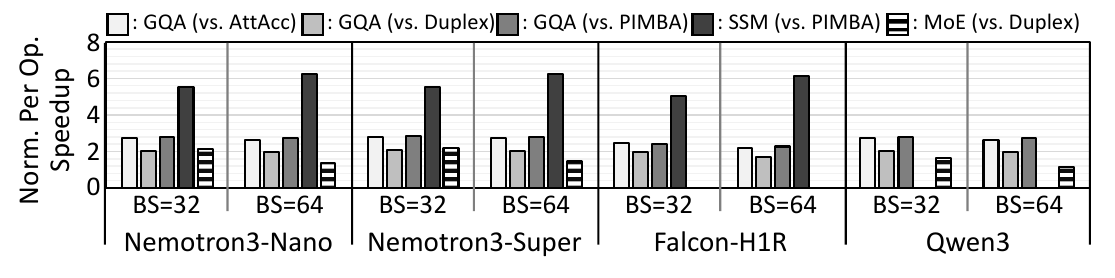}
    \caption {Average decode-phase per-operator speedup of NOVA across sequence lengths.}
    \label{fig12}
\end{figure}

\begin{figure}[t]
    \centering
    \includegraphics[width=1\columnwidth]{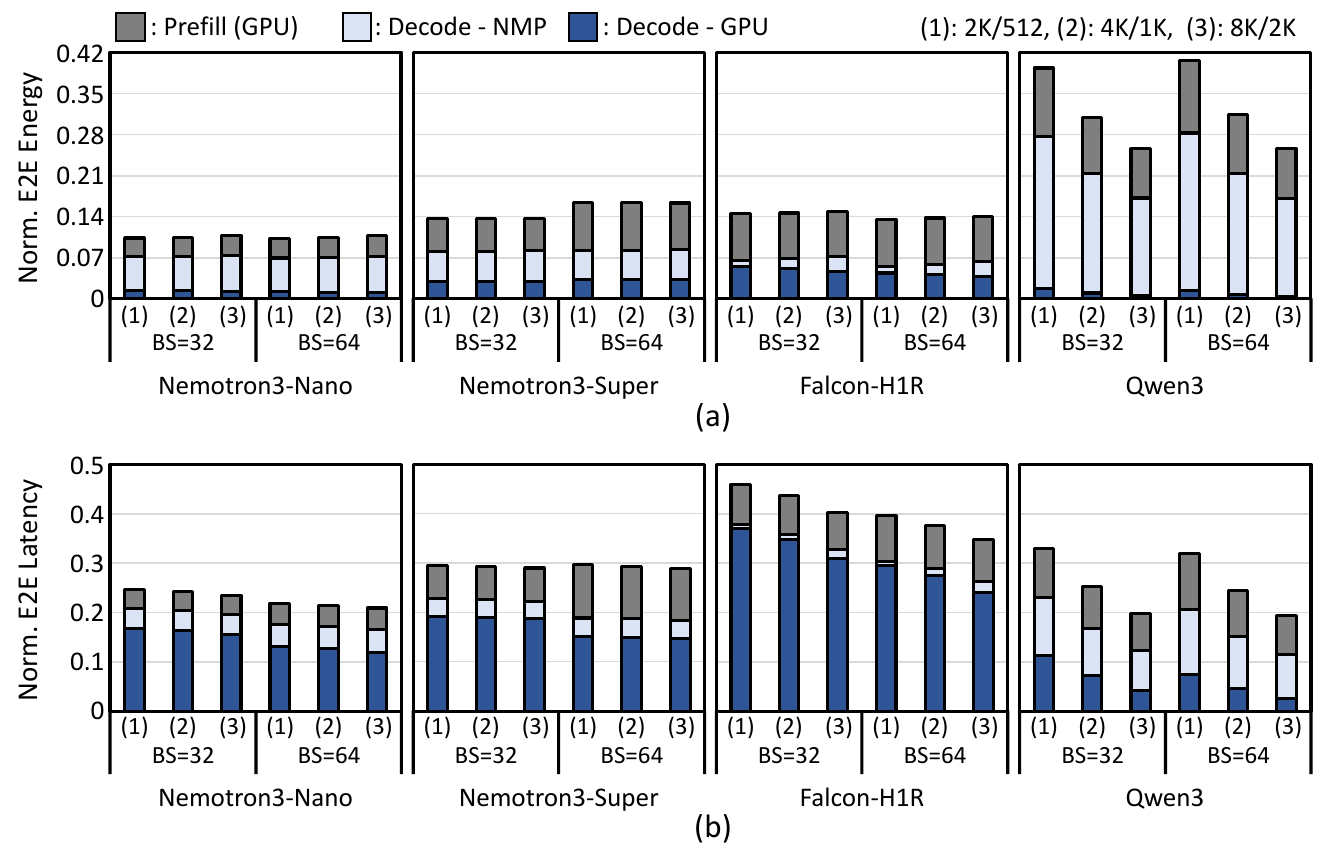}
    \caption {Breakdown of normalized E2E (a) energy and (b) latency under NOVA. Each bar is normalized to its own GPU baseline.}
    \label{fig13}
\end{figure}

\begin{figure}[t]
    \centering
    \includegraphics[width=1\columnwidth]{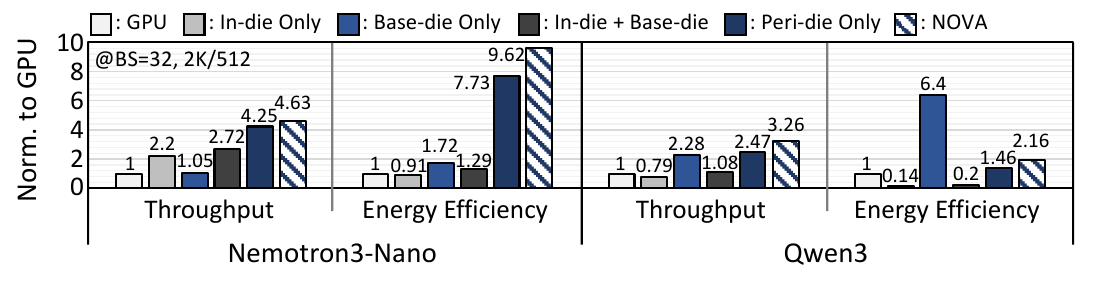}
    \caption {Ablation study of NOVA's NMP tiers.}
    \label{fig14}
\end{figure}

\noindent
\textbf{Per-Operator Speedup.}
Fig. 12 shows the per-operator speedup of NOVA's two-tier NMP architecture over the baselines during the decode phase. The experimental setup is identical to that of Fig. 11, except that the results are averaged across sequence lengths. For GQA, NOVA achieves average speedups of 2.61$\times$, 1.96$\times$, and 2.67$\times$ over AttAcc, Duplex, and PIMBA, respectively, across all models and batch sizes. PIMBA exhibits the lowest performance because, unlike AttAcc, it offloads the aggregation and softmax operations to the GPU. Duplex performs relatively well owing to its base-die design targeting the mid-range Op/B regime, but it still falls short of NOVA. For SSM, NOVA achieves an average speedup of 5.78$\times$ over PIMBA, as the high design flexibility of its logic-process-based peri-die provides both higher compute throughput and higher memory bandwidth than PIMBA. Finally, for MoE, NOVA distributes hot and cold experts to tier-2 and tier-1 at a 1:9 ratio and processes them in parallel, achieving an average speedup of 1.64$\times$ over Duplex.

\noindent
\textbf{Breakdown Analysis.} 
Figs. 13(a) and (b) decompose NOVA's normalized E2E energy and latency against the GPU baseline into three components: the prefill phase executed on the GPU, the decode operations executed on the NMP, and the decode operations remaining on the GPU. Because all prefill and decode phase operations, except for the decode GQA/SSM/MoE operations accelerated by NOVA, are executed on the GPU, their absolute costs are identical to those of the GPU baseline. Furthermore, as each bar is normalized to its respective model's baseline, direct comparisons of bar heights across different models are invalid.

In terms of energy consumption, the NMP accounts for approximately 80\% of the decode phase energy in Nemotron3-Nano, whereas this proportion decreases to about 60\% in Nemotron3-Super. This reduction occurs because Super's larger active parameter size (12B) increases the volume of projection computations remaining on the GPU, and the model's scale incurs additional communication overhead in a dual-GPU configuration. Falcon-H1R exhibits the smallest NMP proportion, as it employs a dense FFN instead of an MoE, meaning the entire FFN is processed on the GPU. Conversely, all 48 layers of Qwen3 consist of GQA and MoE. Combined with the additional activation energy incurred when processing cold experts via all-bank activation in Tier-1, the NMP accounts for the vast majority of the decode phase energy consumption in Qwen3.

For latency, these component proportions are inverted. Because NOVA substantially accelerates the offloaded operations, the processing time of the unaccelerated operations remaining on the GPU becomes relatively more prominent. Specifically, in hybrid models, the input and output projections of GQA/SSM are not targeted for offloading, resulting in a larger share of residual GPU computations. This proportion is most pronounced in Falcon-H1R, where even the dense FFN remains on the GPU. By contrast, because Qwen3 lacks SSMs and its MoE is also processed on the NMP, the NMP accounts for a relatively large proportion of the latency as well.




\subsection{Area Overhead}
The total area overhead incurred by the addition of compute units and TSVs in the NOVA architecture is 5.55 mm$^2$, which accounts for merely a 3.94\% increase relative to the standard HBM4 chip size of 140.8 mm$^2$~\cite{hbm4_isscc}. Specifically, doubling the number of TSVs from 2,048 to 4,096 adds an area of 5.45 mm$^2$ according to~\cite{duplex}, while the BGPU introduces an overhead of 0.1 mm$^2$. The physical area of the BAPU compute unit is 0.127 mm$^2$ per bank. Given the HBM4 chip size, a $4F^2$ DRAM cell density of approximately 78\%~\cite{4f2_isscc}, and the area occupied by TSVs, the available area per bank is estimated to be 0.36 mm$^2$. Consequently, the BAPU occupies about 35\% of this per-bank area. Considering that the POC structure provides about 28\% free space beneath the cell array~\cite{10631320}, and that the DRAM process has lower area efficiency compared to the logic process~\cite{attacc,pimba}, the BAPU can be effectively hidden under the cell array, resulting in zero effective footprint overhead. Furthermore, since the base-die is fabricated using an advanced 4 nm logic process~\cite{hbm4_isscc}, it inherently possesses ample internal free space~\cite{stratum}. Therefore, the 17.7 mm$^2$ required for the 64 Base-PUs—corresponding to approximately 12.6\% of the total HBM4 chip area—is seamlessly integrated into the base-die without incurring any additional area overhead.

\subsection{Discussion}
\noindent
\textbf{Ablation Study.}
To separate the architectural gains of the 2-tier NMP from the technology gains of the logic-process POC structure, Fig. 14 compares five NMP variants under an identical system configuration. In-die only and peri-die only execute all three operations (GQA, SSM, and MoE) on a single tier, where peri-die only is NOVA's POC-based Tier-1 and in-die only is a conventional DRAM-process in-die NMP configured identically to PIMBA (16 BF16 GEMV units with EWM/EWA per bank, 256b granularity). Base-die only uses NOVA's Tier-2 and offloads SSM to the GPU, while in-die + base-die and NOVA execute SSM on the Tier-1 and GQA/MoE in cross-tier parallel mode. All base-die-equipped variants use 4096 TSVs as in NOVA. In-die + base-die thus represents the best 2-tier NMP achievable on a conventional HBM4 stack, so its gap from NOVA isolates the technology-driven gain, while the single-tier gaps isolate the architectural gain of cross-tier parallelism.
Peri-die only consistently outperforms in-die only (e.g., 4.25$\times$ vs. 2.2$\times$ throughput on Nemotron3-Nano), and NOVA outperforms in-die + base-die by 1.7$\times$ and 3.0$\times$ on Nemotron3-Nano and Qwen3, respectively, confirming that the logic-process POC peri-die is the key enabler. On Qwen3, in-die only falls below the GPU baseline (0.79$\times$, 0.14$\times$) because reading massive MoE expert weights at 256b granularity incurs excessive activation commands, while base-die only achieves the highest energy efficiency (6.4$\times$) by avoiding all-bank activation energy. The cross-tier modes trade some of this efficiency for the highest throughput, but NOVA can select its Tier-2-only mode when energy is prioritized. On Nemotron3-Nano, abundant SSM layers let NOVA improve throughput and energy efficiency simultaneously.

\begin{figure}[t]
    \centering
    \includegraphics[width=1\columnwidth]{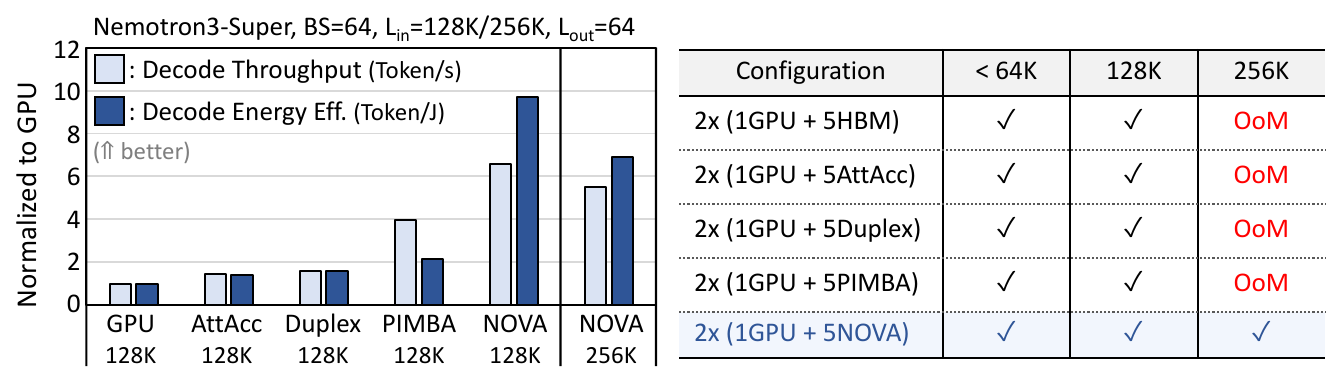}
    \caption {Decode-phase throughput/energy efficiency (left) and capacity feasibility across context lengths (right). NOVA is the only design that runs at 256K without OoM.}
    \label{fig15}
\end{figure}
\begin{figure}[t]
    \centering
    \includegraphics[width=1\columnwidth]{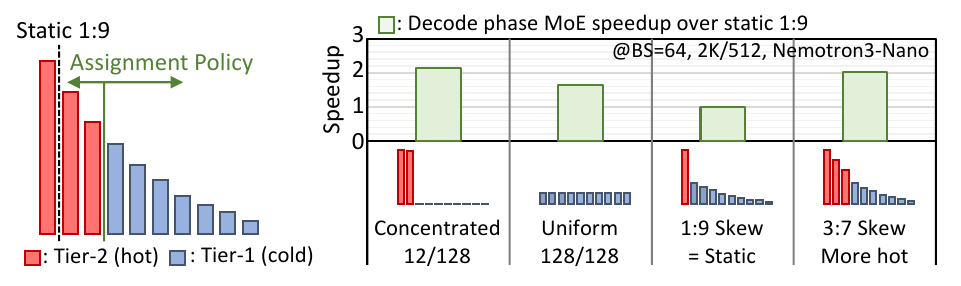}
    \caption {Adaptive tier assignment policy and its speedup over the static 1:9 split.}
    \label{fig16}
\end{figure}

\noindent
\textbf{Capacity-Enabled Feasibility for Long Contexts.}
We evaluate in Fig. 15 the benefit of NOVA's 2$\times$ memory density under an identical system configuration. First, the improved memory density determines the feasibility of long-context inference. At an input sequence length of 256K, the GPU and all prior NMP baselines fail to run due to out-of-memory (OoM), whereas only NOVA can perform inference under the same configuration. In other words, the contribution of the ${4F^2}$ cell does not directly translate into a percent performance gain, but rather extends the workload regime that is infeasible under conventional configurations into a feasible one. Since this means that the same capacity can be met with fewer resources, it is also advantageous in terms of cost.

\noindent
\textbf{Data Assignment Policy.}
All results in Figs.~11--15 use the static 1:9 hot/cold split (see Fig. 4(b)), with hotness profiled offline. A natural concern is robustness to routing distributions that shift across prompts, layers, or time. Because NOVA's mapping is tier-agnostic, the hot/cold split is a compute-assignment decision rather than a data-placement decision, and can be changed at any decode step at zero migration cost, unlike swap-based expert tiering. We formalize the assignment as follows. The attainable throughput of tier $k$ at intensity $I$ is $T_k(I) = \min(C_k, I B_k)$ with $(C_1, B_1) = (32\,\text{TFLOPS}, 16\,\text{TB/s})$ and $(C_2, B_2) = (64\,\text{TFLOPS}, 4\,\text{TB/s})$; pCH interleaving halves Tier-1 to $(16, 8)$ during cross-tier execution ($T_1'(I)$). Splitting work so that both tiers finish together assigns a fraction $\alpha^* = T_1'(I) / (T_1'(I) + T_2(I))$ to Tier-1, and the three modes in Fig. 8(c) are examples of this rule. For MoE, the per-expert intensity equals its routed token count ($I_e \approx n_e$), known from the router output; experts sorted by $n_e$ are split at the boundary $k^*$ where the two tiers' completion times balance ($\sum_{e \le k^*} W_e / T_2 \approx \sum_{e > k^*} W_e / T_1'$), found by binary search at negligible cost. Fig. 16 shows that this adaptive policy matches the static 1:9 split at its design point and yields up to 2.1$\times$ speedup as the routing distribution deviates (concentrated, uniform, and 3:7-skew cases).

\noindent
\textbf{Serving Dynamics.}
NOVA's mapping naturally extends to continuous batching. Because KV and state partitions are managed via fixed address arithmetic, a request joining or leaving the batch simply allocates or frees its slots without requiring any data relocation. Furthermore, the tier-assignment policy dynamically rebalances the workload at every decode step using the current router output and effective batch size. Following the prefill phase on the xPU, new requests join at the subsequent iteration boundary, with their KV data written through the conventional 256b path, completely independent of the BAPU compute path.

%% file: Outline/7_related_works.tex
\section{Related Work}\label{related_work} 
\noindent
\textbf{Hybrid Transformer-Mamba Model Accelerators.}
 HLX~\cite{HLX} is the first unified accelerator for hybrid Transformer-Mamba models, proposing pipelined dataflows (PipeFlash and PipeSSD) and a reconfigurable core (URSC) to address low compute utilization caused by inter-operation dependencies and excessive memory traffic in FlashAttention and state space duality (SSD) kernels. PIMBA~\cite{pimba} accelerates the state update operation in post-Transformer LLMs such as Mamba-2 using in-die NMP with access interleaving and MX8 quantization. However, HLX focuses exclusively on prefill-phase attention and SSD acceleration without MoE support, while PIMBA targets only attention and state update without addressing the group structure of GQA or expert routing in MoE. Consequently, no existing accelerator provides unified support for all three core operations required by state-of-the-art hybrid LLMs.



%% file: Outline/8_conclusion.tex
\section{Conclusion}\label{conclusion} 
This paper presented NOVA, a technology-architecture co-designed NMP system for hybrid LLMs that simultaneously overcomes the Technology and Architecture Walls. By combining the $4F^2$ VCT cell with a POC structure, NOVA doubles memory density at iso-area, and by repurposing the reclaimed logic-process peri-die into a 2-tier NMP, it covers the wide Op/B spectrum of GQA, SSM, and MoE in parallel across the peri-die and base-die without any memory capacity loss. Our evaluation shows that NOVA delivers 4.5$\times$ higher throughput, 69.8\% lower E2E latency, and 5$\times$ better energy efficiency over the GPU baseline at only 3.9\% area overhead.
